\documentclass[12pt]{iopart}
\usepackage{graphicx}

\def\be{\begin{eqnarray}}
\def\ee{\end{eqnarray}}

\usepackage{ulem}
\usepackage{xcolor}

\def\PL{P(p_R)}

\begin{document}

\title[]{Effects of power-law correlated disorders in XXZ spin chain:
Many-body localized to thermal phase transition and its critical regime}

\author{Takahiro Orito\textsuperscript{1,2},
Yoshihito Kuno\textsuperscript{3,4,*},
Ikuo Ichinose\textsuperscript{1}}

\address{$^1$ Department of Applied Physics, Nagoya Institute of Technology, Nagoya, 466-8555, Japan
\\
$^2$ Department of Quantum Matter, AdSM, Hiroshima University, Hiroshima, 739-853, Japan
\\
$^3$ Department of Physics, Graduate School of Science, Kyoto University, 
Kyoto, 606-8502, Japan
\\
$^4$ Department of Physics, University of Tsukuba, Tsukuba, Ibaraki 305-8571, Japan
\\}
\ead{* kuno421yk@gmail.com}
\vspace{10pt}

\begin{abstract}
We study a canonical many-body-localized (MBL) system with power-law-correlated disorders:
$s=\frac{1}{2}$ spin chain in a random magnetic field.
The power-law-correlated disorder can control the critical regime between the MBL and thermal (ergodic) phases by varying its exponent, and it let us investigate the MBL transitions in detail.
Static-eigenstate and dynamic properties of MBL are studied by numerical methods for systems with various
long-range correlations. By using energy-resolved
distribution of the participation ratio (PR) and calculating some physical quantities related to localization length,
we show that the MBL transition exhibits certain universal behavior. We also investigate entanglement properties for the static and dynamics system. These studies elucidate the impact of power-law-correlated disorders in the canonical MBL system.
\end{abstract}

%
%
%
\maketitle
%
%

\section{Introduction}

Many-body localization (MBL) attracts a lot of attentions and interests in condensed matter and quantum information physics these days~\cite{Nandkishore,Abanin,Alet}.
Recent theoretical studies have developed novel points of view of MBL such as entanglement dynamics, 
thermalization properties, and the relationship to quantum integrable systems, as a counter example
of the eigenstate thermalization hypothesis (ETH). 
Development of numerical simulation techniques plays an important role for such trends.
Various isolated quantum systems with inter-particle interactions have been constructed in experiments
on ultra-cold atoms, and they `quantum simulate' MBL phenomena by controlling strength of 
quasi-periodic disorders~\cite{Schreiber,Choi,Lukin,Rispoli}. 
Controllable disorders and interactions between particles have the potential ability to generate various 
novel localization phenomena, which have not been observed in solid state materials. 

In this work, we study effects of correlated disorders in a typical
MBL system, i.e., anti-ferromagnetic $s=\frac{1}{2}$ spin chain in 
a random external magnetic field.
In particular, we focus on effects of disorders with power-law correlations, 
which are feasible in recent experiments. 
[For explicit expressions of the random variables, see Eqs.~(\ref{disorder_correlation}) and 
(\ref{eta2}).]
In the recent studies on MBL, systems with long-range, long-range random, and power-law long-range interactions have been extensively studied by numerical methods~\cite{Deng,Schiffer,Sierant}, and also with 
long-range hopping \cite{Modak}, 
but research on long-range correlated disorders has been still lacking in the study of MBL. 
For Anderson localization, such power-law disorders {\it in non-interacting systems} have been 
extensively studied so far, in particular, from the view point of the localization length and
phase diagram~ \cite{de,Evers,Takeda,Kaya,Shima,DosSantos,Croy,Izrailev,Tessieri,Gurevich,Lugan}. 
However, the extensive study from the modern view point such as entanglement properties, 
localization dynamics and thermalization properties is lacking. 
Therefore in this work, we shall investigate the effects of the power-law disorders
in a systematic way from the above mentioned point of view. 

In the present system, there exists another free parameter besides the disorder strength, i.e., 
the exponent of the power-law.
Motivation for the present work stems from the expectation that study of
the MBL systems under power-law disorders reveal localization nature
that is difficult to be observed by study of 
the systems with the simple short-range on-site random disorders. 
In fact, as we show in the rest of the paper, the critical disorder strength of the MBL to thermal phase transition 
changes its value depending on the power-law exponent of the disorder-correlations.
Furthermore, the critical regime between the MBL and ergodic phases, which exists in finite
size systems~\cite{Khemani}, can be controlled by the exponent.
Time-evolution of the system also changes depending on the exponent.
{\it Careful investigation of these behaviors of the systems reveals that there exists certain
universality for the static-eigenstate and dynamical MBL phase transitions,} although
difference between the two transitions was recognized in the early study of 
the time evolution of the entanglement entropy~\cite{Bardarson}.
For the above study, energy-resolved
distribution of participation ratio (localization length (LL)) plays an essentially important role.

Here, we should remark on the current status of the theoretical study on one-dimensional (1D)
quantum systems with a random-quench disorder such as a quantum spin chain in a
random external magnetic field, as a controversial discussion has 
emerged \cite{suntajs,sierant,abanin,panda}.
Main controversial discussion concerns the question whether the existence of MBL phase transitions can be 
concluded from numerical data for finite-size systems, in particular, in the thermodynamic limit. 
Very recently for time evolution of quantum quench of a 1D fermion system in a random
potential (a Heisenberg spin chain), $\ln\ln t$ evolution of the number entropy was observed for a strong disorder case \cite{keifer,unanyan,max}.
This result indicates the instability of the MBL state observed in finite systems to an ergodic state.
These discussions are ongoing without reaching a final conclusion yet. 
Also, large quantum spin chains and its MBL phase transition have been studied \cite{Chanda}. 
Nevertheless, we shall study an extended system in this work in order to get new insights on MBL.

This paper is organized as follows. 
In Sec.~2, we introduce the target MBL model, the anti-ferromagnetic $S={1 \over 2}$ spin
chain in the random magnetic field with the long-range correlations.
There, we explain physical meaning of the lower-law correlation of the disorder, i.e., it is nothing but
a superposition of disorders with various correlation lengths.
In Sec.~3, we explain the methods to generate the power-law correlated random variables
by making use of the Fourier filtering method (FFM). 
We carefully examine random variables generated by the FFM to find that they display
the desired correlation.
In Sec.~4, non-interacting systems are studied by measuring various quantities, such as the participation ration (PR) and subsystem-size scaling of the entanglement entropy (SSEE).
The problem of how nature of Anderson localization changes due to the correlated random
magnetic field is carefully investigated.
To this end, we calculate energy-resolved distribution of PR, which is closely related with 
the LL as we explain.
Findings in Sec.~4 form basis for study of MBL in Secs.~5 and 6.
Section 5 is devoted to the study of the interacting case.
Various quantities, such as the multi-fractal analysis~\cite{Mace,Yucheng,Luitz2} and 
energy level-spacing ratio (LSR),  are investigated numerically and obtain the phase diagram, in which
a critical regime exists in a finite parameter region.
To understand the critical regime, we calculate the entanglement entropy (EE) and 
the standard deviation of the EE (SDEE) as a function of the disorder strength.
Data obtained for different values of the exponent exhibit certain unexpected behavior.
To understand these numerical results, we explore the distribution of PR corresponding to
relevant parameter regions.
Comparison of the distribution of PR between the non-interacting and interacting systems reveals the origin 
of these unexpected results.
By this observation, we find that there exists `novel universality' for the MBL phase transition.
In Sec.~6, the dynamics of the EE and other related physical quantities are investigated
numerically.
We find that the time evolution of the system exhibits characteristic behavior in the ETH, critical and
MBL regimes.
In particular even in the critical regime, physical quantities evolve with a different power as a function of time 
depending on the exponent of the correlated disorder.
Section 7 is devoted to conclusion and discussion.


\section{Model}

In this work, we consider one of the typical canonical models of MBL, 
$S={1 \over 2}$ XXZ spin model, 
Hamiltonian of which is given by,
\begin{eqnarray}
H_S=&&\sum_{j}\frac{J_{xy}}{2}(S^{+}_{j}S^{-}_{j+1}+S^{-}_{j}S^{+}_{j+1})+J_z S^{z}_{j}S^{z}_{j+1},
\label{XXZmodel}
\end{eqnarray}
where, $S^{+(-)}_{j}$ is a raising (lowering) spin operator, $S^{z}_{j}$ is $z$-component spin operator, 
and $J_{xy}$ and $J_{z}$ are exchange coupling and z-component Ising coupling, respectively. 
Since the $J_{z}$-term acts as an interaction in the Jordan-Wigner fermion picture of
the system [Eq.~(\ref{XXZmodel})], the model is expected to exhibit MBL in the presence of disorders, e.g., a random external magnetic field. 

In this paper, we consider the following random magnetic field as a disorder,
\begin{eqnarray}
H_d=\sum_{j}\eta_{j}S^{z}_{j},
\label{disorder}
\end{eqnarray}
where $\{\eta_{j}\}$ are random variables and have the following specific power-law correlation,
\begin{eqnarray}
\langle \eta_{j}\eta_{j+\ell}\rangle_{\rm ans}\propto  (1+\ell^2)^{-\gamma/2}.
\label{disorder_correlation}
\end{eqnarray}
In Eq.~(\ref{disorder_correlation}), 
$\langle \cdots\rangle_{\rm ans}$ means ensemble average of the disorder $\{\eta_{i}\}$, 
and $\gamma$ is a power-law exponent, which takes various values in the following study.
The method of generating the correlated disorder $\{\eta_{j}\}$ will be explained in Sec.~3. 
The correlation in Eq.~(\ref{disorder_correlation}) reduces to the genuine power-law correlation
 $\sim \ell^{-\gamma}$ for $\ell\gg 1$. 
The merit of the form of Eq.~(\ref{disorder_correlation}) is that 
the singularity for $\ell\to 0$ in the genuine power-law correlation is safely avoided. 
Although the disorders of Eq.~(\ref{disorder_correlation}) slightly deviate from the genuine 
power-law correlation $\ell^{-\gamma}$, they are well suited for studying localization.
Some previous works studied effects of this type of disorder for Anderson 
localization~\cite{Takeda,Kaya,Croy,DosSantos}. 
The localization properties of the systems depend on the parameter $\gamma$, and 
interesting phenomena have been reported for Anderson localization, 
e.g., the violation of Harris criterion \cite{Shima}, the presence of a localization-delocalization phase transition, etc. 
In this paper, we shall investigate how this type of disorder affects the phase diagram of 
the spin model of Eq.~(\ref{XXZmodel}).
That is, we focus on how the spatial correlation of 
Eq.~(\ref{disorder_correlation}) influences localization properties  of the system and compare the obtained
results with those of the spin model in the uniformly random external magnetic field, under which the conventional MBL occurs. 

Before going into the numerical setup and practical calculations, let us consider physical meanings
of the long-range correlated disorder; $\langle \eta_{j}\eta_{j+\ell}\rangle_{\rm ans}\propto  |\ell|^{-\gamma}$
for $|\ell | \gg 1$.
In fact, this long-range correlation can be understood as a superposition of various short-range correlations
such as 
\begin{eqnarray}
\int^\infty_0 dg \; g^c e^{-g|\ell|} \propto |\ell|^{-1-c},
\label{correlation}
\end{eqnarray}
where $c$ is a constant.
From Eq.~(\ref{correlation}), it is obvious that {\it long-range correlations are enhanced for 
$\gamma<1 (c<0)$},
whereas for $\gamma>1 (c>0)$, short-range correlations dominate over long-range ones. 
Then for $\gamma<1$, we expect that the system $H_S+H_d$ exhibits localization properties different
from the ordinary case.
In particular, in the case of $J_z=0$,  extended states to emerge for $\gamma<1$
as we verify later on.
Furthermore, we expect that the above long-range correlations of the disorder generate certain 
localization behavior in the interacting systems, which reveals novel critical properties of of the thermal-MBL
transition.
Detailed will be explained after the numerical studies.

With the above mentioned expectation, we consider $0.2\leq \gamma\leq 2$ throughout this paper. 
For the practical numerical study, the XXZ model of Eq.~(\ref{XXZmodel}) is mapped 
into the fermion system through the Jordan-Wigner transformation. 
Throughout this paper, we set $J_{xy}=1$ and consider the half-filled case in the fermion picture, i.e., 
the $\sum_{j}S^{z}_j=0$ sector in the XXZ model.

\section{Random variables with power-law correlations}
\begin{figure*}[t]
\begin{center} 
\includegraphics[width=10cm]{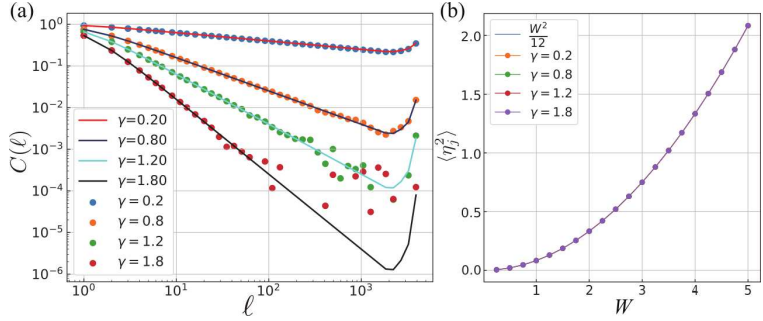}
\end{center} 
\caption{(a) Numerically generated random variables with power-law correlations. 
(b) Variance of the generated power-law random variable as a function of $W$.}
\label{Fig1}
\end{figure*}

In this section, we explain the methods generating random variables $\{\eta_{i}\}$.
In general, the power-law random variables can be produced by employing the FFM, as first discussed in Ref.~\cite{Makse}.
Later, modification of the FFM was developed in Ref.~\cite{Takeda}, and
variance-parameterized versions of the power-law disorder were discussed in Ref.~\cite{Kaya}. 
In this paper, we employ the power-law disorders with the controllable variance parameterized by $W$. 
The value of $W$ is proportional to the strength of the disorder. 
We shall investigate the $W$-dependence of various physical quantities by the numerical study
in later sections. 
Therefore, we employ the FFM with a rescale-variance technique, which was proposed in Ref.~\cite{Kaya}.

To numerically generate $\{\eta_j\}$ for the system with size $L$, 
we start with an ensemble of Gaussian noise
$\{u_{j}\}$ ($j=1,\cdots, L$), whose correlation is simply given by,
\begin{eqnarray}
\langle u_{j}u_{j+\ell}\rangle_{\rm ans}=\delta_{j,j+\ell}.
\label{white_niose}
\end{eqnarray}
From the ensemble $\{u_j\}$, we can construct random variables $\{\bar{\eta}_{j}\}$, which have
the following power-law correlation,
\begin{eqnarray}
C(\ell) = \langle \bar{\eta}_{j}{\bar \eta}_{j+\ell}\rangle_{\rm ans}=(1+\ell^{2})^{-\gamma/2}.
\label{power_niose}
\end{eqnarray}
To generate $\{{\bar \eta}_{j}\}$, we use the Fourier form of the above correlation, i.e.,
\begin{eqnarray}
&\hspace{0.15cm}\mbox{\footnotesize FT}&   \nonumber \\
C(\ell) &\longrightarrow& S(k)= \langle \bar{\eta}(k){\bar \eta}(-k)\rangle_{\rm ans},
\label{power_niose2}
\end{eqnarray}
where FT denotes the Fourier transformation (FT). 
Then the Fourier counterpart of $\{{\bar \eta}_{j}\}$ can be obtained from the Fourier counterpart of 
$\{u_{j}\}$ \cite{Makse},
\begin{eqnarray}
{\bar \eta}(k)=S^{1/2}(k)u(k).
\label{FF}
\end{eqnarray}
Then by applying the inverse Fourier transformation to $\{{\bar \eta}(k)\}$, we obtain $\{{\bar \eta}_{j}\}$ 
with the power-law correlation of Eq.~(\ref{power_niose}). 
Furthermore according to Ref.~\cite{Kaya}, we can add disorder strength to $\{{\bar\eta}_{j}\}$ 
by imposing a normalization condition on the variance of the random variables $\{{\bar \eta}_{j}\}$.
It is achieved by rescaling $\{{\bar \eta}_{j}\}$ as 
\begin{eqnarray}
\eta_{j}=\frac{W}{\sqrt{12 \sigma_{L}}}({\bar \eta}_{j}-\langle {\bar \eta}\rangle_{L}),
\label{rescale}
\end{eqnarray}
where $\sigma_L$, $\langle {\bar \eta}\rangle_{L}$ are the variance and the mean value
of ${\bar \eta}_{j}$'s [$\{{\bar \eta}_{1},\cdots,{\bar \eta}_{L}\}$], respectively.
We obtain a sequence of disorders $\{\eta_{j}\}$ for the target system size. 
Its variance is controlled by $W$ as 
\begin{eqnarray}
\langle \eta^2_j \rangle_{\rm ans}=W^{2}/12, \;\; 
\mbox{with} \;\; \langle \eta_j \rangle_{\rm ans}=0.
\label{eta2}
\end{eqnarray}

For MBL of the Heisenberg spin chain, it is known that the critical strength of the white-noise disorder
depends on the variance of disorder \cite{Janarek}.
We expect that this observation is also applicable to MBL with long-range correlated disorder. 
In this sense, the variance $W$ is one of key parameters that control
localization of the system with the power-law correlated disorder. 
In the later numerical studies, we regard $W$ as the disorder-strength parameter.

We numerically generate variables $\{\eta_{j}\}$ for various values of
$\gamma$ from the white-noise
random variables $\{u_j\}$ generated uniformly.
In Fig.~\ref{Fig1}(a), we show the behavior of the correlation function obtained from
the generated $\{\eta_{j}\}$.
Fig.~\ref{Fig1} (a) shows that $\{\eta_{j}\}$'s with the power-law correlation of 
Eq.~(\ref{power_niose}) are obtained satisfactorily. 
It is noted that for $\ell\gg 1$, the correlation of the numerically obtained $\{\eta_{j}\}$ slightly deviates
from the strict line of Eq.~(\ref{power_niose}), 
but the deviation is less than $\mathcal{O}(10^{-3})$, 
therefore it is negligibly small.
Figure~\ref{Fig1} (b) shows the variance of $\{\eta_{j}\}$'s as a function of $W$.
We observe the good agreement between the numerical results and analytic expression in Eq.~(\ref{eta2}).

\section{XY model: $J_z=0$ case}

Although our target systems are interacting ones, 
we first study effects of the disorder $\{\eta_{j}\}$ for the non-interacting case $J_{z}=0$
in Eq~(\ref{XXZmodel}) and we set $J_{xy}=1$ as the energy unit. 
To investigate the localization properties of single-particle states in the system
$H_S|_{J_z=0}+H_d$, we calculate inverse participation ratio (IPR).
The IPR for each eigenstate is defined as
\begin{eqnarray}
({\rm IPR})_{n}=\sum_{j}|\langle j|\psi_{n}\rangle|^{4},
\end{eqnarray} 
where $|j\rangle$ is the localized single-particle state at site $j$,
and $|\psi_{n}\rangle$ is $n$-th single particle eigenstate. 
For localized states, IPR is close to unity, whereas IPR $\ll 1$ for extended states.
More precisely, if the state $|\psi_n\rangle$ has a non-vanising weight at finite $N_s$ sites uniformly,
$|\langle j |\psi_n\rangle|^2={1 \over N_s}$ for these sites and otherwise zero.
Then, (IPR)$_n={1 \over N_s}$~\cite{Takaishi}.
Likewise for the state with exponential decay $|\langle j |\psi_n\rangle|^2
\propto\exp (-j/\xi)$, 
\begin{eqnarray}
{\rm (IPR)}_n=(\tanh (1/2\xi))^2/\tanh (1/\xi).
\end{eqnarray}
For $\xi \to 0$, (IPR)$_n \to 1$, whereas for $\xi \to$ large, (IPR)$_n \to 1/(4\xi)$.
Thererfore, the participation ration (PR)$_n=1/\mbox{(IPR)}_n$ can be regarded the LL.
 
Here, we should mention 
that the previous works on some correlated disorders have obtained an
interesting observation from IPR indicating the existence of critical regime~\cite{Croy}, 
although the different parameter regime of $\gamma$ from ours was investigated there. 
We focus on the existence of extended states and their location in the present system,
and calculate the participation ratio (PR) for the whole energy eigenstates.

\begin{figure*}[t]
\begin{center} 
\includegraphics[width=12cm]{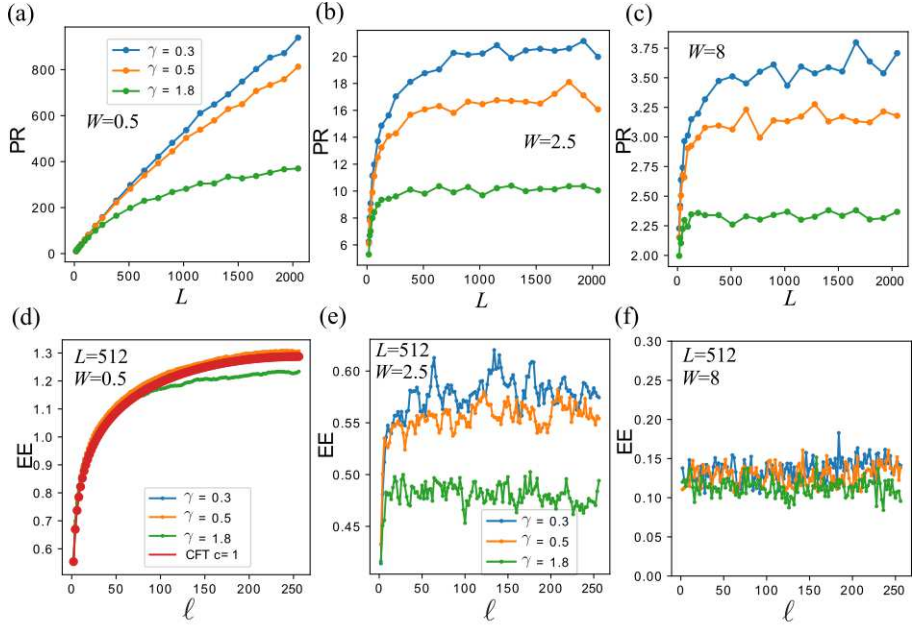}
\end{center} 
\caption{System-size scaling of ${\rm PR}$ for $W=0.5$ (a), $W=2.5$ (b), and $W=8$ (c).
Subsystem-size scaling of the half-filled groundstate EE for $W=0.5$ (d), $W=2.5$ (e), and $W=8$ (f).
Both PR and EE exhibit characteristic behaviors in the delocalized, critical and MBL states.
}
\label{Fig2}
\end{figure*}

In practical calculation, we qualitatively characterize the extended and localized states
by investigating the system-size dependence of the PR.
In the recent study of an extended Aubry-Andre model~\cite{Liu}, the scaling analysis of the PR 
was employed to distinguish localized regime.
Our results of the system-size scaling of the PR are shown in Fig.~\ref{Fig2} (a)-(c) 
for $N_d=600-800$ realizations of $\{\eta_j\}$.
For the case of $W=0.5$ (Fig.~\ref{Fig2} (a)), the PR for $\gamma=0.3$ and $0.5$ linearly
increases with the system size $L$, as a hallmark of the extended phase.
However, for $\gamma=1.8$, the PR tends to saturate for $L>500$.
As we discussed in Sec. 3, the system with $\gamma>1$ is expected to have only localized states,
and the result in Fig.~\ref{Fig2} (a) seems to confirm this expectation.
For $W=2.5$ and $8$ (Fig.~\ref{Fig2} (b) and (c)), the PR for $\gamma=0.3$ and $0.5$ still exhibits
small but finite increase even for $L\sim 2000$.
This result indicates that a finite portion of energy eigenstates is extended in these parameter regimes.
We dare to say that this is a somewhat unexpected result.
We do not think that the above behavior of PR indicates strict suppression of the localized phase.

As the system-size dependence of the PR exhibits rather clear signals of localization and delocalization,
it is interesting to investigate the SSEE of the half-filled groundstate. 
The subsystem size is denoted by $\ell$($<L= 512$).
It is observed that the SSEE well captures the qualitative scaling behavior of the extended and 
localization states \cite{Mondragon,Mondragon2,Peschel,Peschel2}, although sometimes strict scaling 
behavior is not obtained.
In particular for extended state, the SSEE is expected to obey the following CFT scaling 
law~\cite{Calabrese},
\begin{eqnarray}
S_{\rm CFT}(\ell)=\frac{c}{6}\log[(L/\pi)\sin(\pi\ell/L)]+s_0, 
\label{SCFT}
\end{eqnarray}
where $c=1$, $s_0=S_{\rm data}(\ell_0)-\frac{c}{6}\log[(L/\pi)\sin(\pi\ell_0/L)]$, and $\ell_0$ is 
the minimum size of the subsystem.
Although $S_{\rm CFT}$ in Eq.~(\ref{SCFT}) was originally proposed for the critical regime in the thermodynamic 
limit~\cite{Calabrese}, we observe that it quantifies extended states by comparing it with
numerically obtained SSEE for the finite size systems. 

Figure \ref{Fig2} (d)-(f) are the results of the numerical SSEE for various $(\gamma,W)$, 
where we set $L=512$ and $N_d=1\times 10^3$.
In Fig.~\ref{Fig2} (d), typical results of the extended state are obtained
for $\gamma=0.3, \; 0.5$, whereas a small but finite deviation from Eq.~(\ref{SCFT}) is 
observed for $\gamma=1.8$.
For the `critical regime' shown in Fig.~\ref{Fig2}(e), the SSEE slightly increases 
as $\ell$ is increased~\cite{Liu_comp}, but the scaling of the SSEE does not satisfy the CFT scaling 
nor the area law (where SSEE $\sim$ constant).
For a fixed $W$, the smaller $\gamma$ exhibits more increase of the SSEE as a function of $\ell$. 
This implies that the weak power-law decay of the disorder correlation (larger long-range correlation)
enhances the increase of the SSEE, as we expected.  
On the other hand, the localization regime as shown in Fig~\ref{Fig2} (f) exhibits no area law 
scaling, i.e., the SSEE hardly increases as increasing $\ell$. 
This is nothing but the behavior of the localized state.

From the above calculations of the SSEE, we expect that $W=2.5$ is in the critical regime
from the extended to localized phases for $\gamma<1$.
In other words, the localized phase is realized for $W > 2.5\sim 3.0$.
In order to verify this expectation, we calculate distribution of PR=$p_R$
for various values of $W$ for $\gamma=0.2$ and $2.0$.
The energy-resolved distribution of $p_R$, $\PL$, plays an important role in the rest of the present paper.
We calculate $\PL$ as follows.
For each disorder realization, we introduce normalized energy $\epsilon$ defined by 
$\epsilon=(E-E_{\rm min})/(E_{\rm max}-E_{\rm min})$, where $E_{\rm max}$ and $E_{\rm min}$ are
the largest and lowest energy eigenvalues, respectively.
The whole energy spectrum, $\epsilon \in [0,1]$, is divided into nine energy sectors (windows) 
in the descendent order and each sector contains the same number of states.
PR is calculated for 50 eigenstates residing in each sector of the normalized energy, 
and then the PR distribution for each energy sector is obtained by averaging over 2000 disorder realizations.

\begin{figure*}[t]
\begin{center} 
\includegraphics[width=14cm]{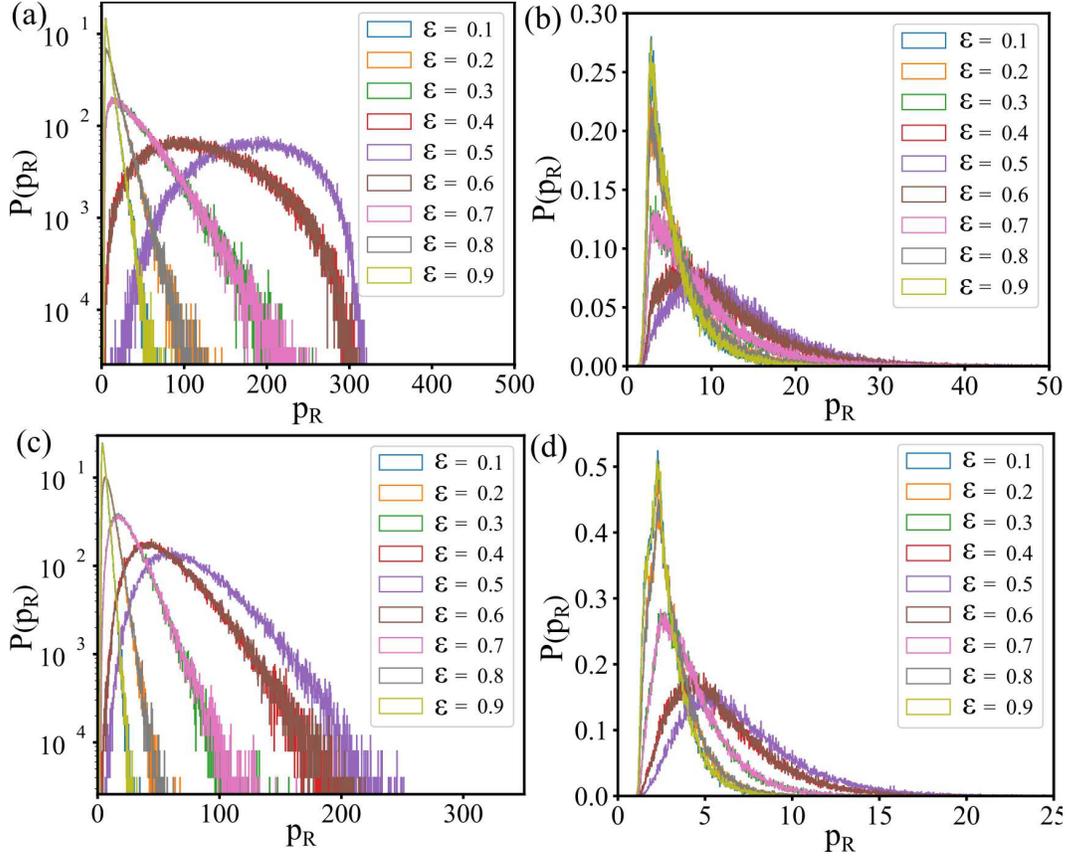}
\end{center} 
\caption{(a) $\PL$ for $(\gamma, W)= (0.3, 1)$. The states in the band center have rather long LL.
(b) $\PL$ for  for $(\gamma, W)= (0.3, 3)$. LLs of all energy sectors are small.
(c) $\PL$ for  for $(\gamma, W)= (1.8, 1)$. 
LLs are considerably shorter that those of $(\gamma, W)= (0.3, 1)$, in particular,
in the vicinity of the band center.
(d) $\PL$ for  for $(\gamma, W)= (1.8, 3)$. LLs of all energy sectors are small.
}
\label{PL1}
\end{figure*}

In Fig.~\ref{PL1}, we show $\PL$ for $W=1$ and $W=3$.
For $(\gamma, W)=(0.3, 1)$, 
$\PL$ shows that the states are extended, in particular, in the vicinity of the band center. 
For $(\gamma, W)=(1.8, 1)$, $p_R$'s are considerably smaller than those of (a).
$p_R$'s for  $(\gamma, W)=(0.3, 3)$ and $(\gamma, W)=(1.8, 3)$ are small in all energy sectors.
We think that the observation via $\PL$ confirms the above conclusion obtained by the SSEE.

The result obtained in the present work should be compared with that of the previous study
for a closely related model.
In Ref.~\cite{de}, it was observed that a regime of extended states exists only for $\gamma<-1$.
We think that this discrepancy stems from the normalization of the random variables $\{\eta_j\}$
in Ref.~\cite{de}.
In fact, the normalization of the random on-site energy is rescaled there by the factor depending
on the system, and the system corresponds to the case of $W=\sqrt{12}$.
As our calculation in the above shows, extended regimes do not exists for $\gamma>0$ with $W>3$,
and therefore our result is consistent with that in Ref.~\cite{de}.

One may think that
the interaction by the $J_z$-term tends to destroy localized states, as the phase coherence
of single-body wave function is destroyed by the interactions, and scattering between
states enhances the entanglement entropy.
On the other hand, the $J_z$--interaction works as a repulsion and therefore single-particle
states tend to get separated with each other.
How the LLs of the many-body systems change is a nontrivial problem.
One may think that the physical picture starting with the single-particle wave functions holds and is useful
for understanding ETH-MBL transition.
In fact, this expectation is supported by the works on topological states in $s={1 \over 2}$ spin chains,
which show the utility of the single-particle picture (the XY model) for understanding edge modes
in the XXZ model~\cite{Hu,Hu2,Orito}.
In the following section, we shall study the ETH-MBL transition of the present system.


\section{XXZ model: $J_z\neq 0$ case}

\begin{figure*}[h]
\begin{center} 
\includegraphics[width=10cm]{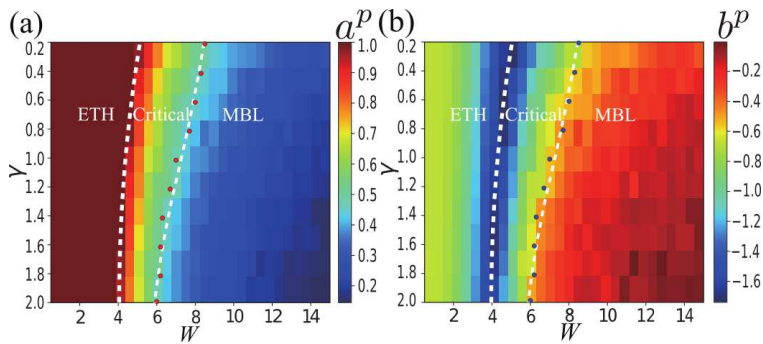}
\end{center} 
\caption{(a) $a^{p}$-distribution in the multifractal analysis, and phase diagram in 
the $(W,\gamma)$-plain. 
The phase boundary between ETH and 
critical phase is determined by the minimum of $b^{p}$. 
The phase boundary between critical and MBL phases is determined by the finite-size scaling 
of the LSR and EE using data of various system sizes. 
The red dots are the transition points, which we numerically observed.
(b) Detailed behavior of $a^{p}$ and $b^{p}$ for typical $\gamma$'s.}
\label{Fig4}
\end{figure*}

Let us turn to the $J_{z}\neq 0$ case, i.e., the interacting case. 
Here, we consider the case with $J_{z}=1$, 
noticing that this case does not have the anti-ferromagnetic order [density-wave phase in the fermionic picture].
As shown in the previous section, the extended and localized phases 
form in the non-interacting case due to the disorders with the long-range correlations.
This is in sharp contrast to the standard model in which the random magnetic field is short-ranged
and all states are localized.
For a finite $J_{z}$, how the phases change is an interesting problem. 
In particular for small $\gamma$ (long-range power-law decay) and moderate $W$, 
we are interested whether the observed critical state is enhanced or not by the interaction.
Furthermore, as the parameter $\gamma$ can control the location of the MBL-thermal
phase transition and also the range of $W$ corresponding to the critical regime, we expect to
obtain new insight into the MBL phase transition by studying the present model with various $\gamma$'s.
As we explain later on, the distribution of PR, $\PL$, plays an important role for that study.


\subsection{Phase diagram: Multi-fractal and level-spacing analysises}

We are interested in {\it static MBL of energy eigenstates} in this and subsequent subsections.
{\it Dynamical MBL} will be studied in Sec.~6.
To obtain the phase diagram, we employ the exact diagnalization (ED)~\cite{quspin1,quspin2}.
To study localization of the system in detail, we use the multi-fractal analysis~\cite{Mace,Yucheng} 
for various values of ($\gamma$,$W$).
In this analysis, Hilbert-space dimensional scaling of the participation entropies 
(PE) denoted by $S_q$ is calculated. 
The quantity $S_q$ is defined by $q$-th moment of wave-function coefficient of each eigenstate 
$|\Psi^n\rangle=\sum_k \psi^n_k|k\rangle$,where $|k\rangle$ is computational basis,
and $|\Psi^{n}\rangle$ is the $n$-th many-body eigenstate. 
Then for $|\Psi^{n}\rangle$, $S_q$ is defined by 
\begin{eqnarray}
 S_q^n=\frac{1}{1-q}{\ln}\biggl[\sum_{k=1}^{{\rm D}}|\psi^n_k|^{2q}\biggr], \; \;
 S_q= \sum_{|\Psi^n\rangle} S^n_q,
\end{eqnarray}
where $D$ is the dimension of the Hilbert space for the system size $L$.
We focus on the quantity of $q =2$ in the present study, i.e., $S_2=-{\ln} ({\rm IPR})$.

The multi-fractal behavior is characterized by the fractional dimension $a^p$ and 
logarithmic subleading correlation term $b^p$.
These are obtained by a fitting such 
as~\cite{Yucheng,Luitz,Mace},
\begin{eqnarray}
\bar{S}_2 =a^{p} (\ln D)+b^{p} \ln (\ln D),
\label{defS2}
\end{eqnarray}
where the coefficients $a^{p}$ and $b^{p}$ 
are obtained by using the $S_2$ data calculated for various system sizes. 
The values of $a^{p}$ and $b^{p}$ are known to characterize three regimes:
extended (ETH), critical and MBL \cite{Mace,Yucheng,Luitz2}.
For $a^{p}\approx  1$ and $b^{p}<0$, the system is in the ETH phase, 
for $0<a^{p}<1$ and $b^{p}<0$ the system is in the critical regime and for  $a^{p}\ll 1$ and $b^{p}>0$ 
the system is in the MBL state.

\begin{figure*}[t]
\begin{center} 
\includegraphics[width=15cm]{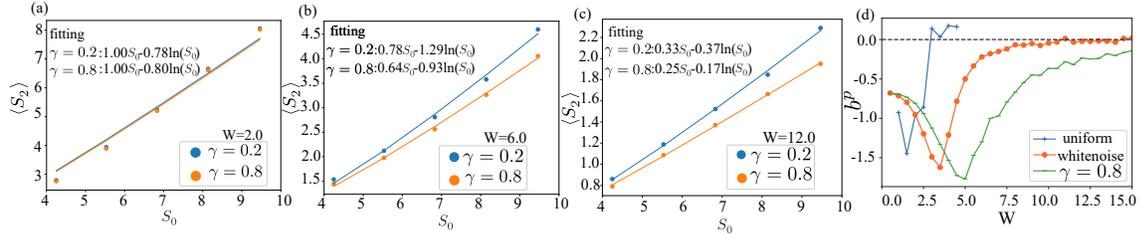}
\end{center} 
\caption{(a)--(c), the data $S_2=-\ln ({\rm IPR})$ vs.
Eq.~(\ref{defS2}) for $W=2.0, \; 6.0$ and $12.0$.
$S_0= \ln D$.
(d) $W$-dependence of $b^{p}$ for various types of disorder. 
Values of $b^{p}$ are estimated by using the data up to $L=16$ system size.}
\label{Fig8}
\end{figure*}

In the practical calculation, in order to perform the multi-fractal analysis,
we first calculate the IPR defined by 
$({\rm IPR})^{n}=\sum_{k}|\langle k|\Psi^{n}\rangle|^4$, 
where 
$|k\rangle$ is the many-body Fock state as reference bases.
We calculate averaged $S_2$, $\langle S_2\rangle$, for the $L=8,10,12,14$ and $16$ systems. 
For the $L<16$ systems, data are obtained by averaging over $12.5\%$ eigenstates of 
the Hilbert-space dimension in the vicinity of the band center and for $10^{2}-10^{4}$ 
disorder realizations. 
For the L=16 system, we use shift invert method and data are obtained by averaging over 250 eigenstates
and for about 600 disorder realizations.
We fit the obtained data of 
$\langle S_2 \rangle$
by Eq.~(\ref{defS2}) as shown in Fig.~\ref{Fig8} (a)--(c),
and obtain the global phase diagram by using obtained
values of $a^{p}$. 
Before going into the detailed investigation 
of the phase diagram, we show some typical fitting data for each regime in Fig.~\ref{Fig8} (a)--(c).
All numerical date can be fitted by Eq.~(\ref{defS2}). 
Therefore, we can extract precise values of $a_p$ and $b_p$. 

As shown in Fig.~\ref{Fig4} (a) and (b), 
the obtained values of $a^{p}$ and $b^{p}$ can be used for identifying the ETH-critical transition 
at least for the system sizes of the present work. 
Notably, we find that the transition regime is located in the minimum of the $b^{p}$ as shown in 
Fig.~\ref{Fig4} (b), and also the transition line is almost independent of $\gamma$.
On the other hand, it is not easy to determine the critical-MBL phase transition line 
only by the results of $a^{p}$ and $b^{p}$. 
For large $W$ regime, 
it is difficult to extract the genuine behavior of $b^{p}$ in our 
system sizes. 
As shown in Fig.~\ref{Fig8} (d), 
the value of $b^{p}$ does not show a clear positive
value, although $b^{p}$ almost approaches to zero.
The zero-approaching behavior of $b^{p}$, however,
gives a possible candidate of the transition line or crossover regime.
Actually, as explained later, the phase boundary between the critical regime and MBL phase can be
extracted by the finite-size scaling analysis of the level-spacing analysis (LSA) and EE.
The ETH-critical phase boundary obtained by the multi-fractal analysis is fairly in good
agreement with that by the LSA.
By using these calculations, the possible phase boundary between critical regime and MBL phase 
is determined as in Fig.~\ref{Fig4} (a). 
It should be emphasized that from our calculation, 
the multi-fractal analysis is efficient to characterize the phase boundary between ETH phase and critical regime. 
As shown in Fig.~\ref{Fig4} (a), we found that the critical regime between the ETH and MBL phases 
is enlarged for small $\gamma$: the long-range power-law disorder.
However, the question whether the critical regime survives 
for the thermodynamic limit is beyond reach of the present work, although some discussion on it
will be given in Sec.~5.2.

Calculations of $b^{p}$ for a different type
of disorder are shown in Fig.~\ref{Fig8} (d).
The behavior of $b^{p}$ in the strong-disorder regime
is different depending on the type of disorder.
In the case of the uniform-disorder ($\eta_i\in [-W,W]$)~\cite{Khemani}, it is
expected that a critical phase disappears and a direct phase transition 
of ETH-MBL takes place in the thermodynamic limit.
On the other hand, for the Gussian-noise disorder such as 
$
\sum_i \eta_i=0, \;  {1 \over L}\sum_i \eta^2_i={W^2 \over 12}
$ 
and long-range correlated disorder ($\gamma=0.8$), the different behavior of $b^p$ appears in Fig.~\ref{Fig8} (d),
the value of $b^p$ does not take a positive value even for the large $W$ as far as our numerical system size.

\begin{figure*}[t]
\begin{center} 
\includegraphics[width=8cm]{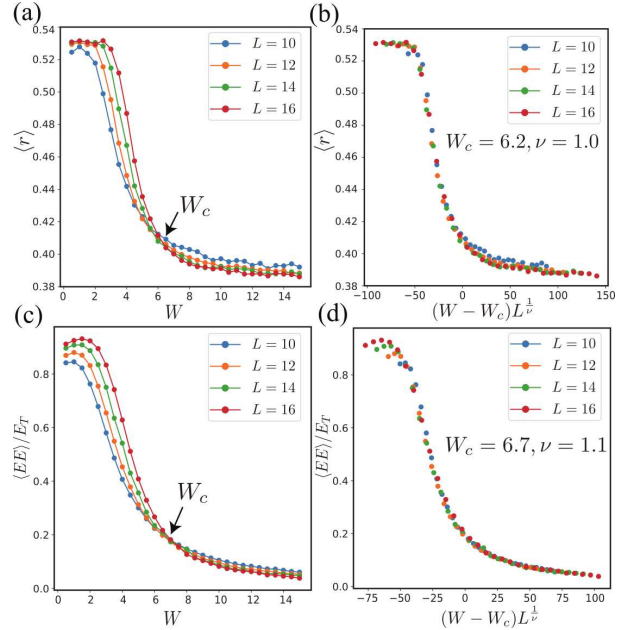}
\end{center} 
\caption{(a) Average LSR $\langle r\rangle$ for $\gamma=1.6$. 
(b) Finite-size scaling of $\langle r\rangle$ for $\gamma=1.6$.
Critical value of $W$ [$W_c$] is estimated as $W_c=6.2$.
(c) Average entanglement entropy $\langle EE\rangle$ for $\gamma=1.6$, 
scaled by the Page value $E_T=0.5(L\ln{2}-1)$ \cite{Page}. 
(d) Finite-size scaling of $\langle EE\rangle$ for $\gamma=1.6$.
}
\label{Fig5}
\end{figure*}

Next, we calculate another quantity to detect the phase boundary between 
the critical regime and MBL phase.
In order to corroborate the phase diagram in Fig.~\ref{Fig4} (a), 
we calculate the average level-spacing ratio (LSR) $\langle r \rangle$~\cite{Janarek,Luitz}.
For calculating the LSR $\langle r \rangle$, we first obtain the spectrum $\{E_{i} \}$ (in the ascending order). 
For each level spacing $\{E_{i} \}$, we define
$r^{k}=[{\rm min}(\delta^{(k)}, \delta^{(k+1)})]/[{\rm max}(\delta^{(k)},\delta^{(k+1)})]$, 
where $\delta^{(k)}=E_{k+1}-E_{k}$.
Value of $\langle r \rangle$ is obtained by averaging over hybrid samples obtained by disorder
realizations and $12.5\%$ eigenstates of the Hilbert space dimension in the vicinity of the band center. 
For the L=16 system, we use shift invert method.
This calculation gives the clear result of $\langle r\rangle$~\cite{Luitz}.  
The value of $\langle r\rangle$ characterizes the ETH and MBL phases;
for the ETH phase, $\langle r\rangle \sim 0.53$ (Gaussian orthogonal ensemble), 
for the MBL, $\langle r \rangle \sim  0.386$ (Poisson random matrix ensemble), 
and the intermediate values of $\langle r\rangle$ indicates the critical regime.
Figure~\ref{Fig5} (a) shows the $W$-dependence of $\langle r\rangle$ for a typical $\gamma$ and 
various system sizes. 
All data move from $\langle r\rangle \sim 0.53$ to $\langle r\rangle \sim 0.386$ 
as $W$ increases. 
Globally, all data exhibit behavior of the ETH-MBL transition. 
Notably, the results obtained for various system sizes shown in Figs.~\ref{Fig5} (a) tend to 
intersect with each other at a single point, $W \sim 6.2 \equiv  W_{c1}$. 
From the above multi-fractal analysis, this
regime corresponds to the critical-MBL transition.
Hence, our results of the LSA imply the presence of a phase
boundary between the critical regime and MBL phase, and the estimated
$W_{c1}$ is a candidate for the
phase transition point separating the critical regime and MBL phase. 
Furthermore, we estimate the critical exponent $\nu$
by using finite-size scaling with respect to $W_{c1}$.
See Fig.~\ref{Fig5}. 
We obtained $\nu=1.0\sim 1.4$, which clearly breaks the Harris criterion, $\nu=2$~\cite{Harris}.
In particular, for larger $\gamma$ (approaching to the white noise) the value of $\nu$ gets close to the value obtained in the conventional MBL phase transition~\cite{Luitz}. 
Here, we should also comment that the phase boundary for large $\gamma$'s obtained
in the present work is consistent with that of the limit $\gamma \to \infty$ , i.e., white-noise disorder,
extensively studied in Ref.~\cite{Janarek}.

In addition, the EE is calculated to complement the LSA.
The half-chain EE, $S_A$, 
is calculated as $S_{A}=-{\rm Tr}[\rho_A \log \rho_A]$, 
where $\rho_A$ is the partial density matrix of the half chain that is obtained from
a many-body eigenstate of the full system.
The system size dependence for $\gamma=1.6$ system is displayed in Fig.~\ref{Fig5} (b).
Here, similarly to the LSR, the calculations of the EE intersect with each other at a single point, 
$W \sim 6.7 \equiv  W_{c2}$, which is very close to the value of the LSR. 
Here, we estimate the critical exponent $\nu$. 
The value is very close to the value of $\nu$ in the LSR.
From above observation, 
we expect that the critical regime--MBL transition is
observed by both the LSR and EE finite-size scaling analysis.

From the results of the multi-fractal analysis, LSR and EE, 
we conclude that in the present power-law disorder system, 
the critical regime separating the ETH and MBL phases exists at least for finite systems that we studied.
In the subsequent section, we study this critical regime and properties of the MBL phase
transition of the static-energy-eigenstates by using the distribution of PR, $\PL$.
As far as we know, this point of view has not to be employed so far for the study of the MBL phase transition.


\begin{figure*}[t]
\begin{center} 
\includegraphics[width=10cm]{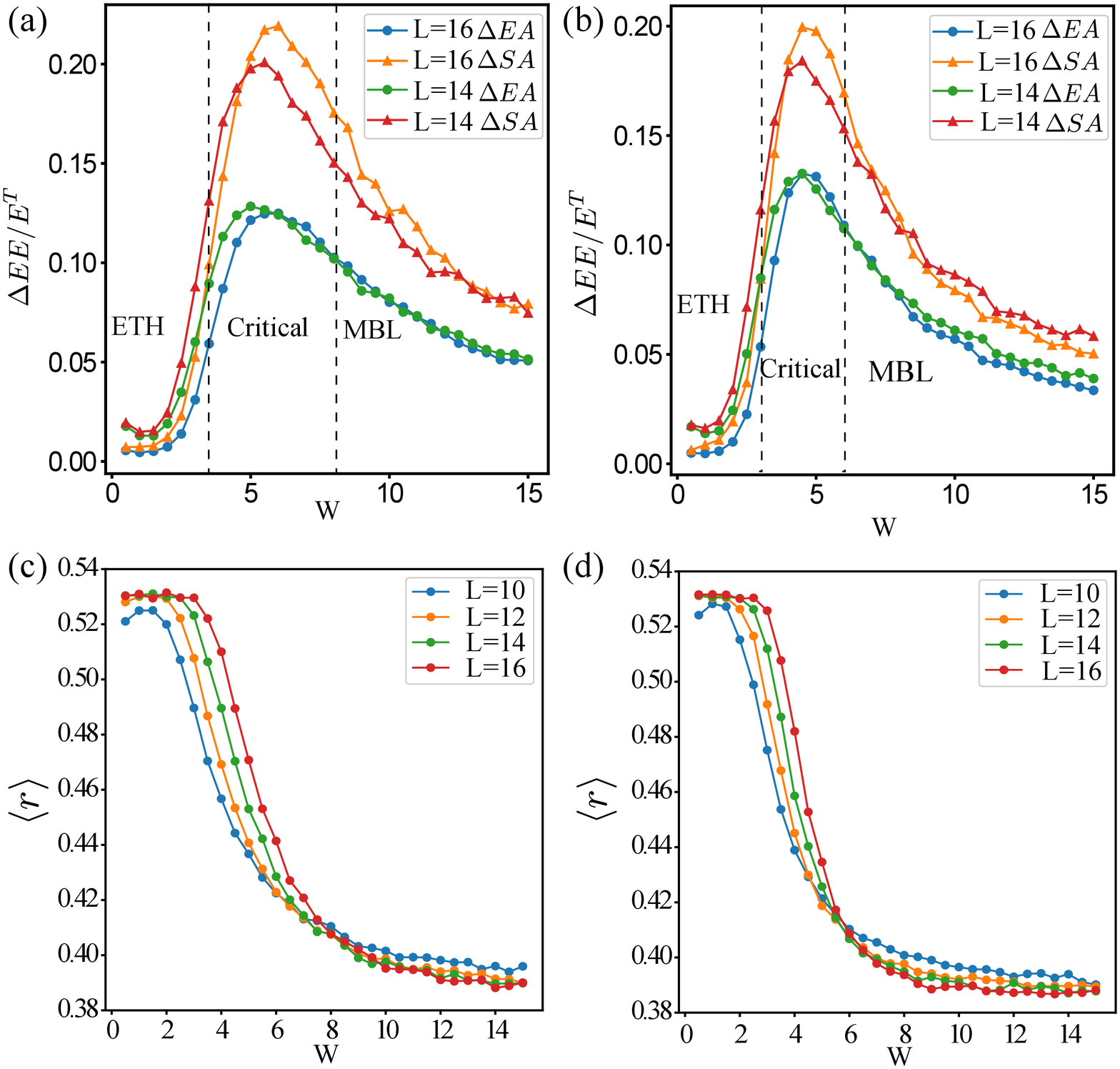} 
\end{center} 
\caption{(a) SDEE for $\gamma=0.2$.
Critical $W$'s obtained by LSR are shown by dashed lines, 
one of which is given by the system-size
independent point, and the other is determined by the plateau of $\langle r \rangle$. 
See Fig.~\ref{SDEE} (c) below.
Region between two critical points corresponds to the critical regime.
(b) SDEE for $\gamma=2.0$.
For larger $\gamma$, the critical regime is smaller.
Values of $\Delta_{\rm EA/SA}$ for $\gamma=0.2$ and $2.0$ are almost the same at
the critical points.
(c) LSR for $\gamma=0.2$. 
$W$ for the crossing is estimated as $W_c=8$.
ETH regime terminates at $W\simeq 3.5$ for $L=16$.
(d) LSR for $\gamma=2.0$. 
$W$ for the crossing is estimated as $W_c=6$.
ETH regime terminates at $W\simeq 3.0$ for $L=16$.
}
\label{SDEE}
\end{figure*}


\subsection{Critical regime and MBL transition: Study by participation-ratio distribution}

\begin{figure*}[h]
\begin{center} 
\includegraphics[width=14cm]{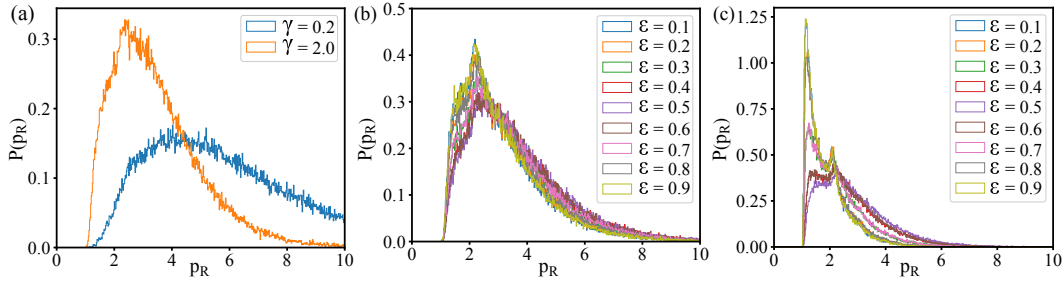}
\end{center} 
\caption{DLL for the non-interacting systems.
(a) $\PL$ in the vicinity of the band center for $W=5$ and $\gamma=0.2$ and $2.0$.
Distributions of the localization length are quite different from each other for  $\gamma=0.2$ and $2.0$,
although the SDEE in Fig.~\ref{SDEE} has a peak at $W\simeq 5$ for both cases.
(b) $\PL$ for $\gamma=0.2$ and $W=8$, in the vicinity of the MBL-critical transition point.
(c) $\PL$ for $\gamma=2.0$ and $W=6$, in the vicinity of the MBL-critical transition point.
$\PL$ for the above two cases exhibit rather different distribution, although they both correspond to
the MBL-critical transition point. 
}
\label{PL0220}
\end{figure*}

In the previous subsection, we obtained the phase diagram of the present model in the $(W-\gamma)$ plane.
We would like to characterize the critical regime in the phase diagram by using specific physical quantity,
i.e., the SDEE.
We define two kinds of SDEE,
which we call sample-to-sample and eigenstate-to eigenstate deviations, respectively.
Definitions of them are given as follows in terms of the EE of state $j$ and disorder realization $i$, $S^j_i$,
\be
&&\langle S\rangle \equiv {1 \over N_SN_E}\sum^{\rm sample}_i
\sum^{\rm state}_j S^j_i,  \nonumber  \\
&&\langle S\rangle_i \equiv {1 \over N_S}\sum^{\rm state}_j S^j_i,  \nonumber
\ee
\be
&&\Delta_{\rm SA}=\Big( {1\over N_SN_E}
\sum^{\rm sample}_i\sum^{\rm state}_j(S^j_i-\langle S\rangle)^2\Big)^{1/2}, \nonumber \\
&&\Delta_{\rm EA}={1\over \sqrt{N_S}N_E}\sum^{\rm sample}_i\Big(\sum^{\rm state}_j
(S^j_i-\langle S\rangle_i)^2\Big)^{1/2},
\ee
where $N_S \ (N_E)$ is the number of disorder samples (eigenstates) used for evaluation, and
$S^j_i$ is the entanglement entropy of eigenstate $j$ in sample $i$.
Usually $\Delta_{\rm SA}>\Delta_{\rm EA}$, and for the case in which
fluctuations across samples are very small,  $\Delta_{\rm SA}\simeq\Delta_{\rm EA}$.

We focus on the cases with $\gamma=0.2$ and $2.0$ in order to see the $\gamma$-dependence of 
the system more clearly.
[The results of of $\gamma=0.4$ and $1.8$ cases are given in appendix B.]
In Fig.~\ref{SDEE}, we display the SDEE and LSR [$\langle r \rangle$].
In both cases with $\gamma=0.2$ and $2.0$, the SDEE exhibits a peak around $W\sim 5$.
On the other hand, the system-size analysis of LSR has a fixed point at $W_c \sim 8$ and $6$
for $\gamma=0.2$ and $2.0$, respectively.
One may wonder what causes this discrepancy between the LSR and SDEE as the both quantities are regarded 
as measures of the ETH-MBL phase transition.
Careful look at the SDEE shows that the peak of the $\gamma=0.2$ is broader than that of 
the $\gamma=2.0$ case.
In fact, this peak originates from the mixing of various eigenstates with various localization lengths;
extended and localized, and therefore the peak of the SDEE identifies the critical regime whose existence was
speculated in the previous subsection.
By using the LSR, we explicitly indicate the ETH, critical and MBL regimes in Fig~\ref{SDEE}.
Then, it is a quite important and interesting question if the critical regime survives 
in the thermodynamic limit $L\to \infty$.
Large system-size calculations are needed to answer this question. 

In order to understand the phase transition out of MBL and properties of the critical regime,
the distribution of PR is quite useful.
We calculate $\PL$ using shift invert method in the vicinity of the band center for 250 eigenstates 
and 160 disorder realization.
In Fig.~\ref{PL0220}, we show $\PL$ for $W=5.0$, $\gamma=0.2$ and $2.0$.
One may expect a similar distribution of PR at $W=5$ for $\gamma=0.2$ and $2.0$ as
$W=5$ is the center of the critical regime for both $\gamma$'s,
but  $\PL$ for $J_z=0$ in Fig.~\ref{PL0220} shows that the PR for $\gamma=0.2$ and $2.0$ 
has a quite different distribution.
This is somewhat unexpected result.
In fact, we expect that states with long LL generate volume entanglement of the system.
Similarly in the vicinity of the MBL-critical phase transition, which is located at $W_c \sim 8$ and $6$
for $\gamma=0.2$ and $2.0$, respectively, $\PL$'s for $J_z=0$
in Fig.~\ref{PL0220} show that the distributions are different with each other. 
The above results of $\PL$ seem rather odd if one expects that wave functions evolve smoothly 
as $J_z$ increases, and the single-particle LLs shed light on the MBL phenomena.

Then, we are interested in the LL of the present interacting system.
Since the straightforward definition of the LL does not exist for the interacting many-body system, 
we employ the methods of the one-particle
density matrix proposed in Ref.~\cite{Bera}, which can extract the LL of a single particle in the system even for interacting systems.
For the present spin system, we introduce the spin-correlation matrix (SCM) such as~\cite{Hopjan},
\be
S^\pm_{ij}= \langle \psi_n |S^+_iS^-_j|\psi_n\rangle.
\label{SCM}
\ee
From the SCM in Eq.~(\ref{SCM}), we obtain the eigenvalues and eigen-vectors such as,
\be
S^\pm |\phi_\alpha\rangle=s_\alpha |\phi_\alpha\rangle,
\label{SCMeigen}
\ee
where eigen-vectors  $|\phi_\alpha\rangle$ are called natural orbits and the eigenvalues, $s_\alpha$,
correspond to their occupations. 
The natural orbital $|\phi_{\alpha}\rangle$ is expected to be related to a localized bit, namely, a localized dressed spin picture \cite{Bera}, the broadening of which can be regarded as the LL.
By using the natural orbitals and their occupations in Eq.~(\ref{SCMeigen}), 
IPR of the present many-body system is defined as follows,
\be
{\rm IPR}_{\rm mb}={1 \over S^z+L/2}\sum^{L}_{\alpha=1} s_\alpha \sum_{j=1}^L
|\phi_\alpha(j)|^4,
\label{IPR2}
\ee
where $S^z={1 \over L}\sum_j \langle S^z_j \rangle$ and $S^z=0$ in the present case.
It is easy to show that the IPR$_{\rm mb}$ is close to unity for localized $|\psi_n\rangle$'s,
whereas  IPR$_{\rm mb}\propto 1/L$ for extended states.
Therefore, the participation ratio
PR$_{\rm mb}=1/\mbox{IPR}_{\rm mb}$ is a typical length of the many-body states in 
the present spin system.

\begin{figure*}[t]
\begin{center} 
\includegraphics[width=10cm]{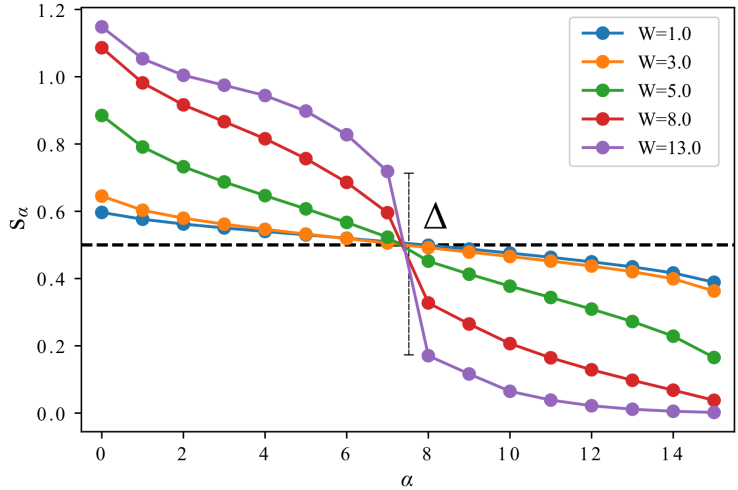} 
\end{center} 
\caption{Occupations $\{s_\alpha\}$ for $\gamma=0.2$ in the descendant order
for various values of $W$.
$\Delta= s_{L/2}-s_{L/2+1}$.
}
\label{salpha}
\end{figure*}

We show our numerical results where we consider the case of $S^z=0$ as before and mostly study the states in the vicinity of the band center.
The typical number of the disorder realization is 160, and we take 250 states.
We first display typical examples of the occupation $\{s_\alpha\}$ in the descendant order in Fig.~\ref{salpha}.
The results show the expected behavior of  $\{s_\alpha\}$.
However compared with the calculations for the standard white-noise system, which was obtained in Ref.~\cite{Hopjan},
the value of $\Delta= s_{L/2}-s_{L/2+1}$ is rather small for large values of $W$.
Anyway, the results in Fig.~\ref{salpha} indicate the existence of the critical region for $W=8.0$ and $5.0$,
but the system for $W=3.0$ is close to that of $W=1.0$, and seems delocalized.

\begin{figure*}[t]
\begin{center} 
\includegraphics[width=10cm]{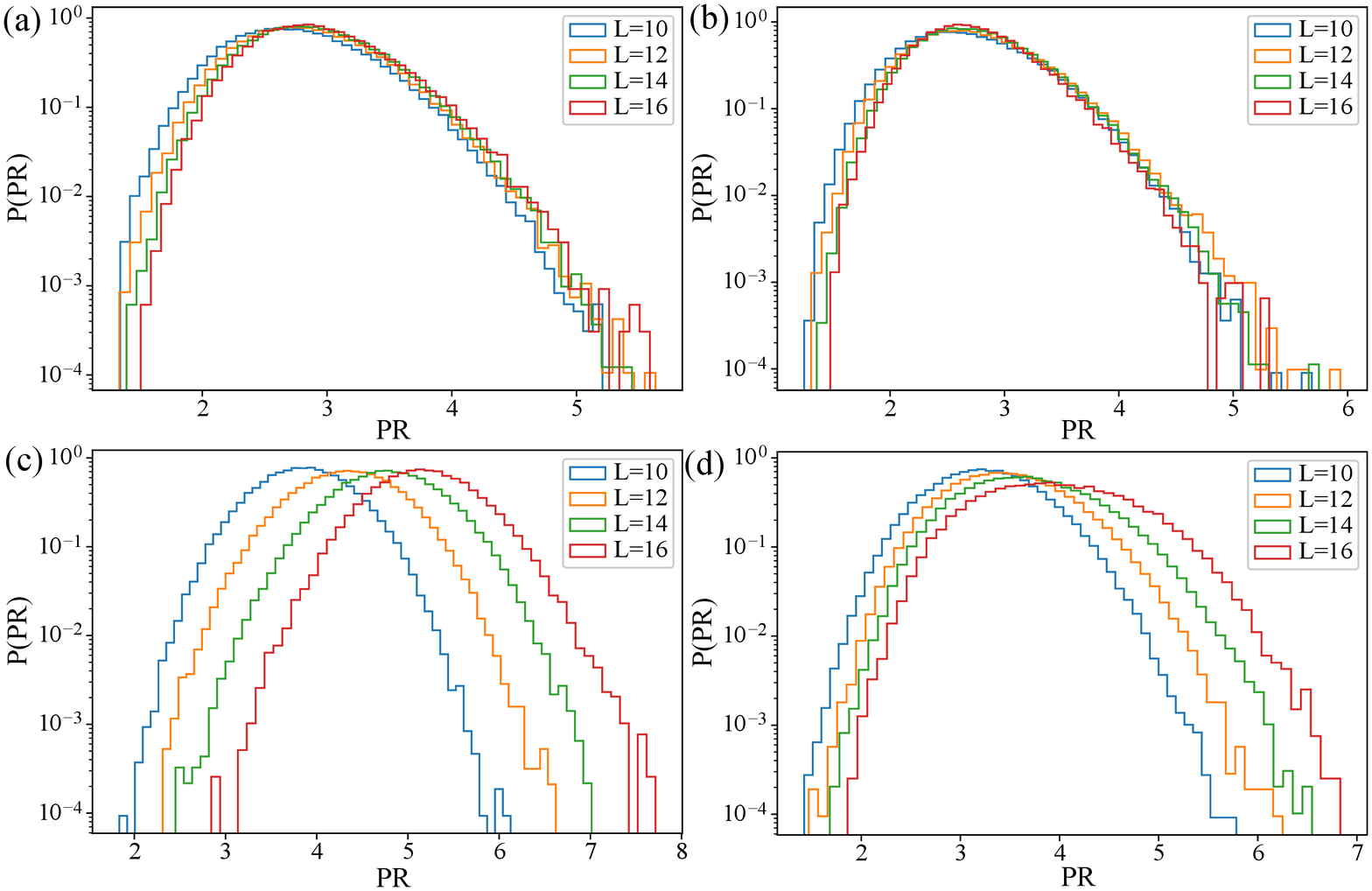} 
\end{center} 
\caption{$\PL$ for $W=5.0$ and $\gamma=0.2 \mbox{(a)}/\gamma=2.0\mbox{(b)}$.
Compared with PR for the non-interacting case in Fig.~\ref{PL0220}, $\PL$'s are rather close
with each other.
$\PL$ exhibits small but finite system-size dependence, which indicates the system is in critical regime
as the other calculations suggest.
(c) $\PL$ for $W=1.0$ and $\gamma=0.2$.
(d)  $\PL$ for $W=3.0$ and $\gamma=0.2$.
Clear system-size dependence is observed for $\PL$ in the extended regime.
}
\label{PRmb1}
\end{figure*}
\begin{figure*}[h]
\begin{center} 
\includegraphics[width=8cm]{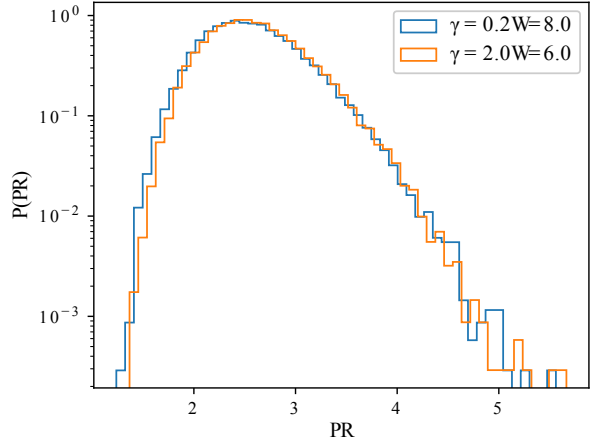} 
\end{center} 
\caption{$\PL$ for $(\gamma, W)=(0.2, 8)$ and $(\gamma, W)=(2.0, 6)$, which correspond to
the MBL-critical transition point.
Obviously, PR$_{\rm mb}$'s are close with each other.
$J_z=1$ and $L=16$.
}
\label{PRmb2}
\end{figure*}

In Fig.~\ref{PRmb1}, we display calculations of PR$_{\rm mb}$ for $W=5.0$ and $\gamma=0.2, \;2.0$.
This parameter corresponds to the critical regime observed by the LSR and SDEE shown in Fig.~\ref{SDEE}.
It is interesting to compare the results in Fig.~\ref{PRmb1} with those of the non-interacting case
in Fig.~\ref{PL0220} (a).
As the definition of PR is different for the interacting and non-interacting cases, the comparison of
the absolute value of PR is meaningless.
However, we find that the PR$_{\rm mb}$ in Fig.~\ref{PRmb1} is rather close with each other for the above parameters,
which is in sharp contrast to $\PL$ in  Fig.~\ref{PL0220} (a).
This contradicts the brief that states evolve smoothly from the Anderson-localized ones
as the interactions are added.
Furthermore, close look at the results in Fig.~\ref{PRmb1} reveals that $\PL$ has a small but
finite system-size dependence, and therefore the states at the above parameters are not fully
localized.
For comparison, we show $\PL$ for $W=1.0$ and $3.0$ with $\gamma=0.2$ in Fig.~\ref{PRmb1} (c) and (d).
$\PL$ tends to shift to large values of $p_R$ as the system size is increased.
This is a typical behavior of $\PL$ in the extended regime.

In Fig.~\ref{PRmb2} , we display $\PL$ for $J_z=1$ with  $(\gamma, W)=(0.2, 8)$ and $(\gamma, W)=(2.0, 6)$, 
which correspond to the MBL-critical transition point for each $\gamma$.
The results show that they are quite different from those of $J_z=0$ in Fig.~\ref{PL0220}, and
interestingly $\PL$'s have almost the same distribution.
This fact obviously indicates the existence of novel universality of the MBL transition as the above two critical
points correspond to quite different parameters.
We found that $\PL$ does not have significant system-size dependences in the localized regime.
Therefore, the above result for $\PL$ indicates the possibility that $\PL$ can be a fingerprint
of the phase transition out of the MBL state.
This observation is one of the most important findings of the present work.
Further study on this universality is certainly desired. 
This a future work.

A few comments are in order. 
The typical length $p_R$ for $J_z=1$ are rather short. 
This behavior obviously supports the local-bit picture of MBL~\cite{Serbyn,Huse,Imbrie}. 
It also indicates that phase transition out of the MBL takes place {\it without divergent LLs.}
This is in sharp contrast to Anderson localization of the single-particle physics.
In other words, the phase transition out of the MBL phase is a phenomenon of inter-particle correlations.
This picture seems to support scenario of sparse backbone of small thermal blocks for 
MBL transition~\cite{Khemani}. 
$\PL$ for $W=5$ exhibits slightly different behavior for $\gamma=0.2$ and $2.0$.
One may wonder if this difference can be observed by measuring certain physical quantities.
This question is answered in the following section. 
$\PL$ exhibits almost the same distribution in the vicinity of the phase transitions for $\gamma=0.2$ and $2.0$.
As the LLs are short there, the long-range properties of the disorder do not influence the LL distribution
although the location of the transition is influenced by the exponent of the correlation.


\section{Dynamics of entanglement entropy and imbalance: $J_z\neq 0$ case}

\begin{figure*}[h]
\begin{center} 
\includegraphics[width=7cm]{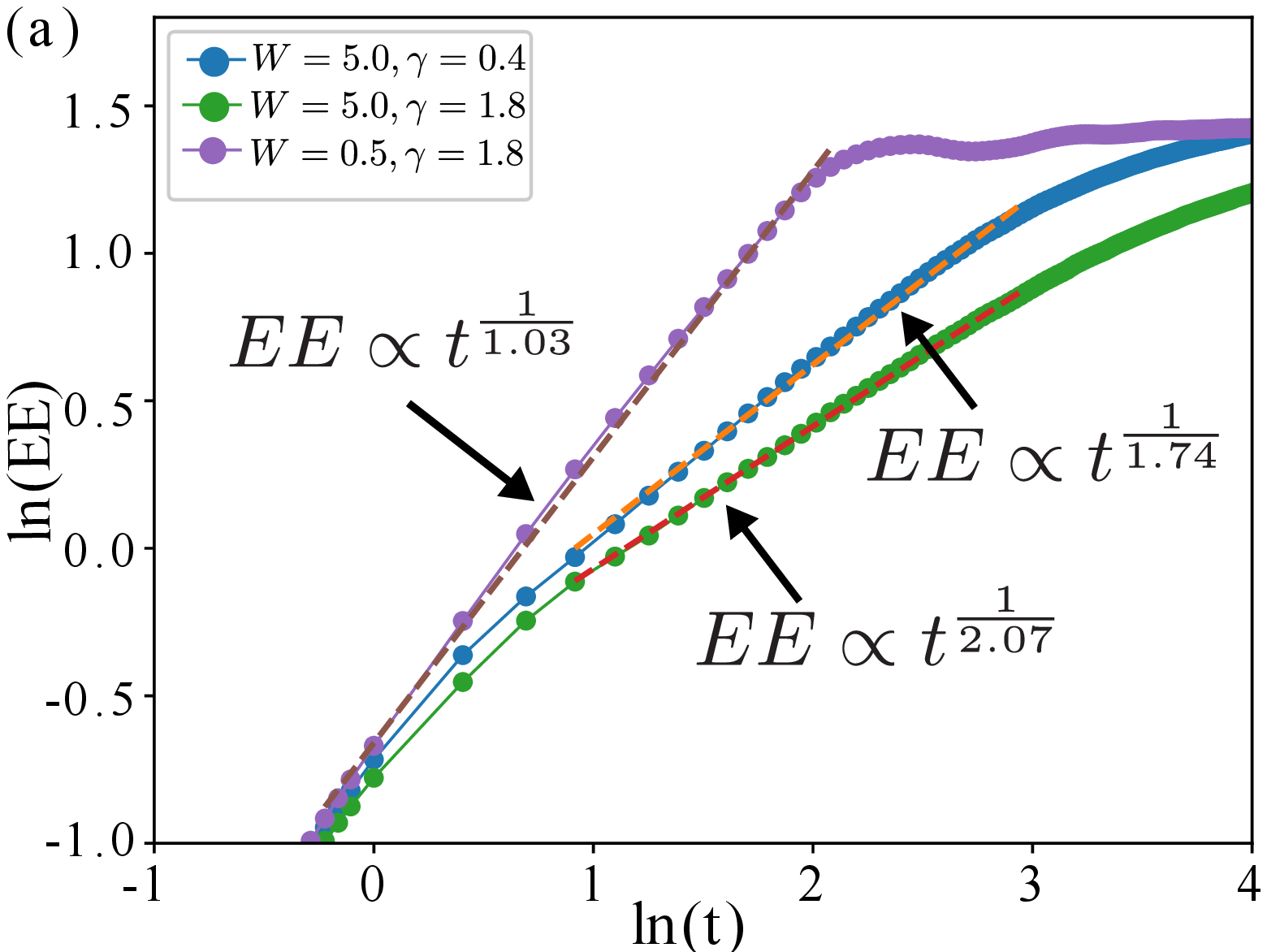} 
\includegraphics[width=6.8cm]{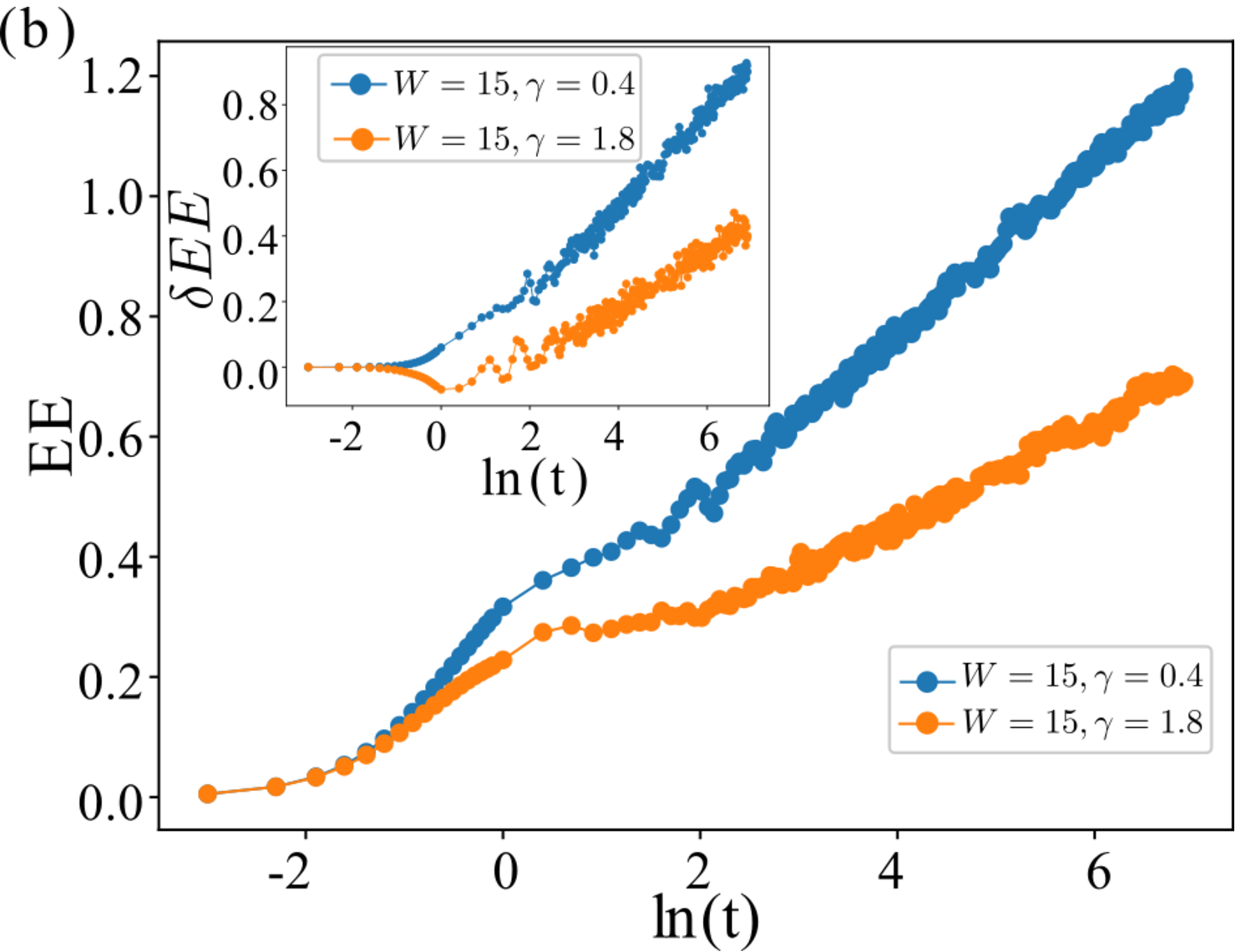}
\end{center} 
\caption{(a)
Time evolution of entanglement entropy (EE)
for points in $(\gamma, W)$ corresponding to the ETH ($W=0.5$)
and critical ($W=5.0$) regimes. 
EE exhibits a power-law behavior.
System size is $L=18$, $J_z=1$ and the open boundary condition is employed. 
Data of $50$ disorder realizations are averaged for each point in $(\gamma, W)$.  
(b) EE for the MBL state ($W=15$) with $J_z=1$ cases. 
EE $\propto \log (t)$ as expected.
Inset displays the difference, $\delta$EE, between
$J_z=0$ and $J_z=1$ cases:
$\delta$EE$=E(J_z=1)-E(J_z=0)$.}
\label{Fig10}
\end{figure*}

In the previous section, we observed that $\PL$ shows the universal behavior
at the phase transition point out of the MBL state, although
the critical regime of static-eigenstate 
exhibits slightly different $\PL$ depending on the value of $\gamma$.
In this section, we shall investigate the dynamics of the EE and imbalance
to see if these quantities exhibit different behavior reflecting $\PL$.
It is recognized that regime of the dynamical MBL phase is generally different from that of the static-eigenstate
MBL~\cite{Chandran1}.
This fact was realized at the very beginning of the measure of the dynamical properties of 
the MBL state \cite{Bardarson}.
The difference between the static-eigenstate and dynamical MBLs is plausible in our picture
via $\PL$ because various states with different energies emerge in the time-evolution of states even though
the value of $W$ is fixed in the evolution.

For the finite-$J_z$ system, the dynamics of the EE is studied with a time-dependent 
many-body wave function of the full system.
In general, time evolution of the EE is used to distinguish the MBL state from
other states such as Anderson localized, ETH phases, as the EE 
exhibits a very slow evolution in the MBL state. 
In particular, if an initial state is a local product state such as the N\'{e}el state, 
the EE changes its value in the time evolution because of 
dephasing effect of the  state~ \cite{Serbyn,Chandran,Huse}. 
It is known that such dephasing is weak in the
MBL state compared with the thermal state, and as a result, 
increase of EE is very slow.

Figures~\ref{Fig10} (a) and (b) exhibit the EE as a function of time, $t$, for various values of $\gamma$ and $W$. 
Here, we employed the N\'{e}el state as an initial state.
In Fig.~\ref{Fig10}, we display the data by the
linear-log as well as log-log plots.
For small $W=0.5$, which corresponds to the ETH state, EE $\propto t$ and it saturates to a finite
value close to unity.
This is nothing but a ballistic evolution of the EE.
On the other hand for the MBL regime ($W=15$), the increase of
the EE is very slow such as
EE $\propto \ln(t)$, which is a hallmark of the MBL state.
We also observe a characteristic $\gamma$-dependence of the time evolution of the EE, i.e..,
for smaller $\gamma$,
system exhibits stronger dephasing in the MBL phase. 
Therefore, the power-law disorder can control the rate of increase of the EE. 
Such control may be possible in recent cold-atom experimental systems \cite{Schreiber,Choi,Lukin,Rispoli}.
Finally for the critical regime with $W=5.0$, the EE displays time evolution that is in-between of 
the ETH and MBL states.
For $\gamma=0.4$ and $1.8$ with $W=5$, the EE has a different power-law as a function
of $t$ depending on the value of $\gamma$, although the systems with $W=5$ correspond to
the center of the critical regime of both $\gamma=0.4$ and $1.8$, as we observed in Sec.~5.
This result seems to indicate that the dynamics of the system depends on $\PL$.

\begin{figure*}[t]
\begin{center} 
\includegraphics[width=7cm]{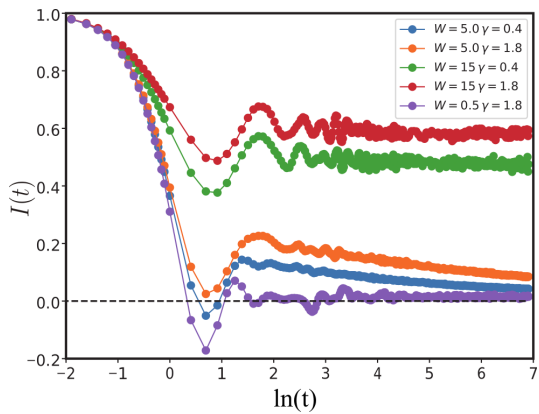}
\end{center} 
\caption{Time evolution of the spin z-component imbalance for ETH, critical and MBL phases.
Each phase exhibits its typical behavior.
}
\label{Fig12}
\end{figure*}

Finally, we study the time evolution of the imbalance of the $z$-component of the system spin,
which is defined as
\begin{eqnarray}
I(t)=(S^{z}_{o}-S^{z}_{e})/(S^{z}_{o}+S^{z}_{e}),
\end{eqnarray}
where $S^{z}_{o}=\sum_{j\in {\rm odd}}S^{z}_{j}$ and $S^{z}_{e}=\sum_{j\in {\rm even}}S^{z}_{j}$.
For the numerical study, we employ the N\'{e}el state as the initial state, 
and we average the calculations over 50 disorder realizations.
Figure.~\ref{Fig12} displays results of the averaged $I(t)$ for typical values $(\gamma,W)$ corresponding to
the ETH, critical and MBL regimes.  
For $W=0.5$, $I(t)$ approaches to the vanishing value after oscillation.
This behavior is expected as the system is in the ETH state.
On the other hand for $W=15.0$, $I(t)$ keeps a finite value as $t \to \infty$
for both $\gamma=0.4$ and $1.8$ cases.
This result obviously indicates that the system is in the MBL state.
The case $W=5.0$ exhibits the behavior of $I(t)$ that is in-between of the ETH and MBL states,
and there exists small but finite difference in $I(t)$ for $\gamma=0.4$ and $1.8$.
All the above calculations of $I(t)$ support the observations obtained so far.
In appendix C, we show calculation of $I(t)$ for a domain-wall initial state to find 
similar behavior of $I(t)$ to the above.



\section{Conclusion and discussion}

In this work, we have systematically investigated effects of power-law correlated disorder 
not only for the non-interacting case but also for the system with the many-body interactions.
We clarified that for both the non-interacting and interacting systems, the critical regime exists 
between the ETH and MBL phases and it is enhanced by the long-range correlations of the disorder.
In particular for the MBL system, we obtained the detailed phase diagram by making use of 
the multi-fractral analysis, LSR and EE calculations. 
In order to understand the phase structure of the systems, 
the distribution of PR is useful.

Then, we studied the critical regime of static-eigenstate MBL by calculating the SDEE,
and compared it with the LSR.
We found that the peak of the SDEE represents the mixing of extended and localized states, which
characterizes the critical regime, and identified the parameter region of the disorder strength
for the critical regime. 
Interestingly, the location of peak of the SDEE is almost the same for various values of $\gamma$,
whereas t$\PL$ in the non-interacting system exhibits rather different 
distribution depending on $\gamma$.  
In order to understand this `discrepancy', we calculated that $\PL$ in the interacting
case with $J_z=1$ and found that $\PL$ exhibits close profile for the above parameters.
Next, we investigated $\PL$ in the vicinity of the transition point out of MBL for $\gamma=0.2$
and $2.0$, and found that it has almost the same distribution $\PL$.
We also found that the LL is rather short at the transition point compared to Anderson localization,
and concluded that the local-bit picture seems correct and MBL is a phenomenon of 
`strongly-correlated' fermions.

Finally, we investigated the dynamics of the EE and imbalance under the power-law disorder.
For the critical and MBL parameter regimes, the their dynamics exhibits different behavior in the
time evolution indicating the existence of a transition (or crossover) between two phases,
although if it survives or not in the thermodynamic limit is a difficult problem.
The time evolution of the EE and imbalance depends on the exponent $\gamma$ in contrast
to the static-eigenstate properties.
We think that this behavior stems from the difference in $\PL$'s for various energies,
besides the band center.
The above result indicates that the power-law disorder has potential ability to control the evolution of the EE.
Such control of the correlations in the disorders is feasible in recent experiments on cold atoms.
Therefore, we expect that physical phenomena originating from the long-range disorders will be
observed by experiments on ultra-cold atoms, trapped ions, etc.

One of the most important results in this paper is that we 
showed the utility of $\PL$, and found the universal properties of the MBL transition
by $\PL$.
Dynamical behavior of the disorder systems also can be understood by means of the PR picture.
As we showed, $\PL$ is quite different in Anderson localization with $J_z=0$
and MBL with $J_z=1$.
Then, it is an interesting problem to see how the above two regimes are connected with each other,
in other words, how the single-particle picture changes by increasing the strength of the interaction.
Concerning to the topological nature of the spin chain,  the utility of the single-particle picture was verified.
Recently, an interesting work toward this direction was given in Refs.\cite{Bera,Hopjan}.
Anyway, these are future works.


\paragraph{Author contributions}
T.O. and Y.K. contributed equally to this work.

\paragraph{Funding information}
Y. K. acknowledges the support of the Grant-in-Aid for JSPS
Fellows (No.17J00486).


\begin{appendix}

\section{Calculations of $b^{p}$ in $\langle S_2 \rangle$ for various types of disorder}

Here, we show the detail calculations of $b^{p}$ in the multi-fractal analysis.
We would like to verify the utility of $b^{p}$'s multi-fractal analysis to determine
the phase diagram in finite systems. 
In fact, we found that the utility depends on the types of disorder.

We calculated for $W$-dependence of $b^{p}$ for the following three cases;
(i) uniform random disorder, $\in [-W,W]$, 
(ii) white noise, $\sum \eta_i=0, \sum {1 \over L}(\eta_i-\langle \eta\rangle)^2={W^2 \over 12}$, and 
(iii) Power law disorder with $\gamma=0.8$.
The results are shown in Fig~.\ref{Fig8}.
In the case of the uniform disorder, the value of $b^{p}$ becomes sufficiently positive for large $W$. 
This indicates that $b^{p}$ correctly characterizes the MBL phase transition even in the system of 
$L=16$.
On the other hand for the cases of both the white noise and power-law disorder, 
$b^{p}$ does not have a positive value in the $L=16$ system size even for large $W$, 
although the value of $b^{p}$ is fairly close to zero for the white noise, 
and takes positive value sometimes. 
These results indicate that calculation of the parameter $b^{p}$ by itself is not sufficient for 
characterizing MBL phase transition at least for the system size $L=16$. 


\section{SDEE of $\gamma=0.4$ and $1.8$ systems}

\begin{figure*}[h]
\begin{center} 
\includegraphics[width=10cm]{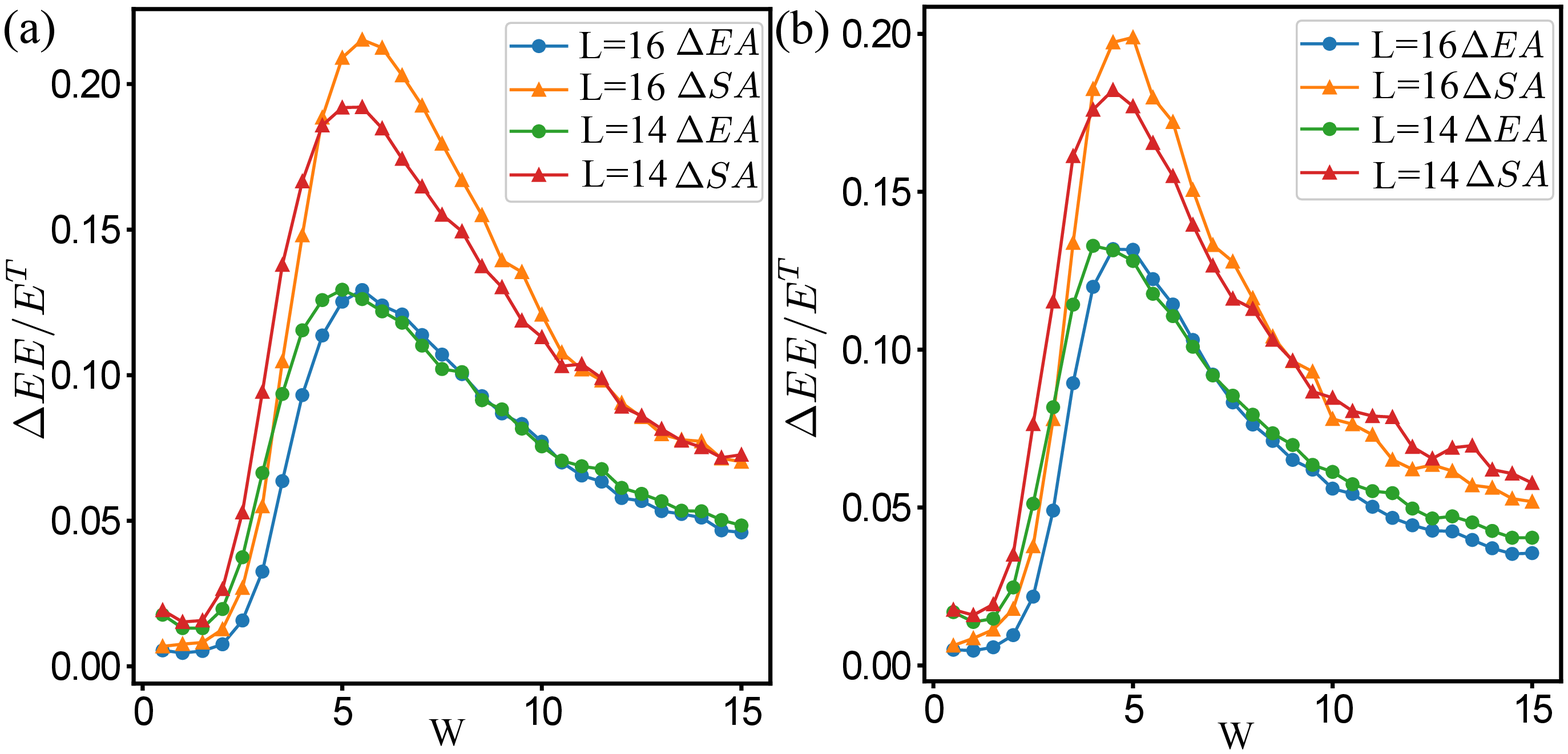}
\end{center} 
\caption{SDEE of $\gamma=0.4$ and $1.8$ systems.
System seize $L=14$ and $16$.
}
\label{SDEE2}
\end{figure*}
Figure~\ref{SDEE2} shows SDEE of $\gamma=0.4$ and $1.8$ systems.
$\Delta_{\rm EA}$ and $\Delta_{\rm SA}$ are defined as explained in the text.
For both $\gamma=0.4$ and $1.8$. the peaks are located $W\simeq 5$, and the maxima of 
two peaks are almost the same.
However, distribuntion of the LL, $\PL$, exhibits somewhat different profile for two cases.


\section{Time evolution of entanglement entropy for a domain wall initial state}
\begin{figure*}[h]
\begin{center} 
\includegraphics[width=6cm]{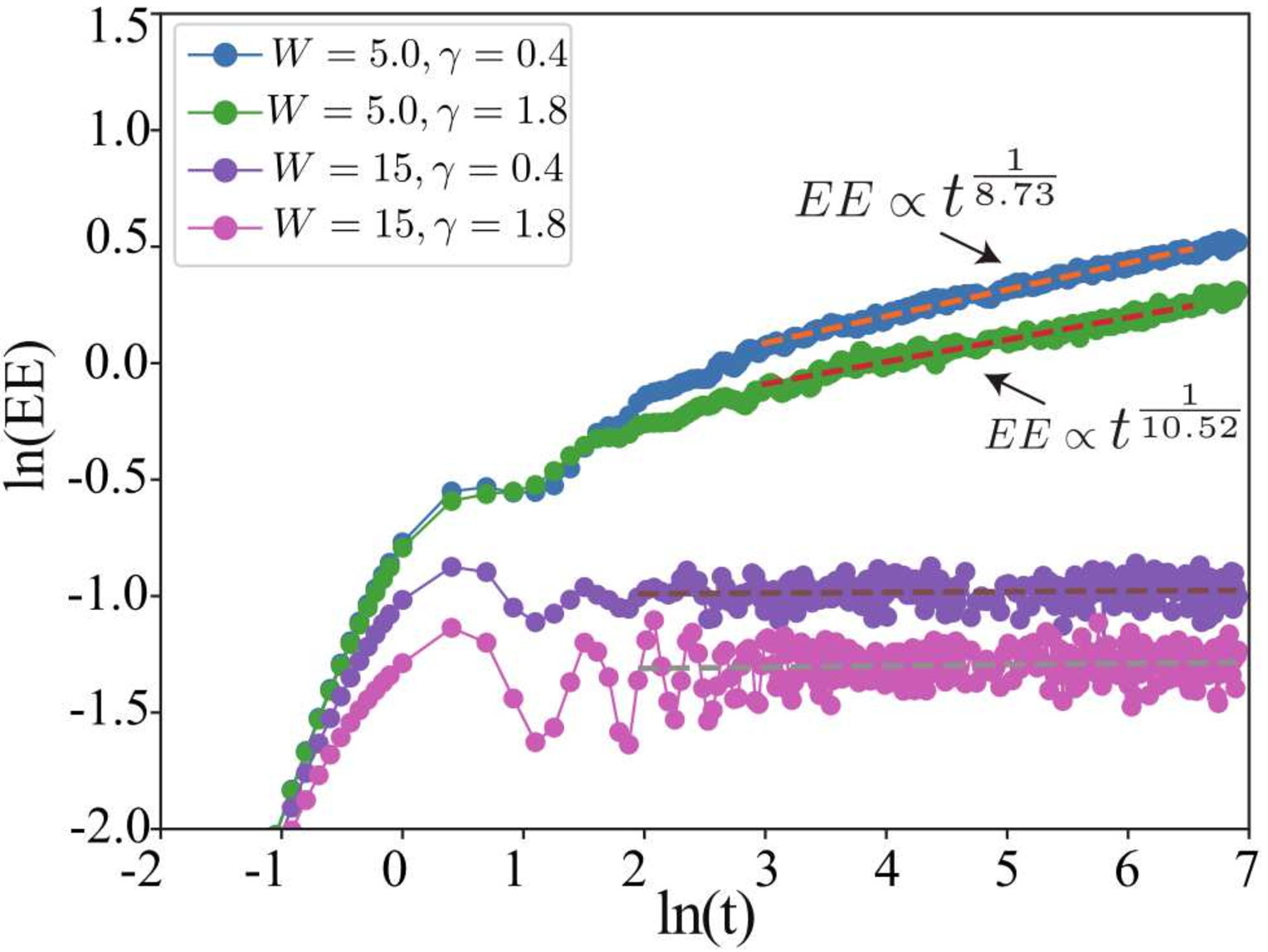}
\end{center} 
\caption{Time evolution of the entanglement  entropy for various $(\gamma, W)$. 
System size $L=18$, and open boundary condition is used. 
Data are obtained by averaging over $50$ disorder realizations.
}
\label{Fig14}
\end{figure*}

We study time evolution of the EE and related physical quantities for a different initial state.
We set a domain wall configuration: up spin are set from $j=1$ to $j=L / 2$ and 
down spin are set from $j=L/2+1$ to $j=L$, therefore the kink is located between 
$L/2$ and $L/2+1$-th sites.
For the critical and MBL phase regimes, the numerical results are shown in Fig.~\ref{Fig14}. 
Here we found that in the critical regime, the EE exhibits power-law  growth in the time evolution and 
the evolution rate is larger for smaller $\gamma$. 
On the other hand in the MBL phase, the EE saturates to a finite values after a finite period, and
the saturating value is larger for smaller $\gamma$.

\end{appendix}




\section*{References} 
\begin{thebibliography}{99}
\bibitem{Nandkishore}
Nandkishore R and Huse D A 2015  
{\it Annu. Rev. Condens. Matter Phys. }{\bf 6} 15 

\bibitem{Abanin}
Abanin D A and Papic Z 2017
{\it Annalen der Physik} {\bf 529} 1700169

\bibitem{Alet}
Alet F and Laflorencie N 2018 {\it Comptes Rendus Physique} {\bf 19} 498 

\bibitem{Schreiber}
Schreiber M, Hodgman S S, Bordia P, Luschen H P, Fischer M H, Vosk R,  Altman E, Schneider U
and Bloch I 2015 {\it Science} {\bf 349} 842 

\bibitem{Choi}
Choi J Y, Hild S, Zeiher J, Schaus P, Rubio-Abadal A, Yefsah T, Khemani V, Huse D A, 
Bloch I and Gross C 2016
{\it Science} {\bf 352}, 1547 

\bibitem{Lukin}
Lukin A, Rispoli M, Schittko R, Tai M E, Kaufman A M,  Choi S, Khemani V, Le\'onard J and 
Greiner M 2019 {\it Science} {\bf 364}, 256

\bibitem{Rispoli}
Rispoli M, Lukin A, Schittko R, Kim S, Tai, M E, Le\'onard J and Greiner M 2019
 {\it Nature} {\bf 573} 385 

\bibitem{Deng}
Deng X, Masella G, Pupillo G, and Santos L 2019
 arXiv:1912.08131 

\bibitem{Schiffer}
Schiffer S,  Wang J, Liu X and Hu H 2019
{\it Phys. Rev.} A {\bf 100} 063619

\bibitem{Sierant}
Sierant P, Biedro\'{n} K, Morigi G, and  Zakrzewski J 2019
{\it  SciPost Phys.} {\bf 7} 008

\bibitem{Modak}
Modak R and Nag T 2020 arXiv:1903.05099 

\bibitem{de}
de Moura F A B F and Lyra M 1998
{\it Phy. Rev. Lett.} {\bf 17} 3735 

\bibitem{Evers}
Evers F and Mirlin A D 2008 
{\it Rev. Mod. Phys.} {\bf 80} 1355 

\bibitem{Takeda}
Takeda K and Ichinose I 2003
{\it Nucl. Phys.} B {\bf 663} 520 

\bibitem{Kaya}
Kaya T 2007
{\it Eur. Phys. J.} B {\bf 55} 49

\bibitem{Shima}
Shima H,  Nomura T, and Nakayama T 2004
{\it Phys. Rev. B} {\bf 70} 075116 

\bibitem{DosSantos}
Dos Santos I F, de Moura F A B F, Lyra M L and Coutinho-Filho M D 2007
{\it J. Phys. Condens. Matter} {\bf 19} 476213 

\bibitem{Croy}
Croy A, Cain P and Schreiber M 2011
{\it Eur. Phys. J.} B {\bf 82} 107 

\bibitem{Izrailev}
Izrailev F M and Krokhin A A 1999
{\it Phys. Rev. Lett. } {\bf 82} 4062

\bibitem{Tessieri}
Tessieri L and Izrailev F M 2001
{\it Phys. Rev. }E {\bf 64} 066120 

\bibitem{Gurevich}
Gurevich E and Kenneth O 2009
{\it Phys. Rev. }A {\bf 79} 063617 

\bibitem{Lugan}
Lugan P, Aspect A, Sanchez-Palencia L, Delande D, Gremaud B, Muller C A and Miniatura C 2009
{\it Phys. Rev.} A {\bf 80} 023605 

\bibitem{Khemani}
Khemani V, Lim S P, Sheng D N and Huse D A 2017
{\it Phys. Rev. }X {\bf 7} 021013

\bibitem{Bardarson}
Bardarson J H, Pollmann F and  Moore J E 2012 
{\it Phys. Rev. Lett. }{\bf 109} 017202

\bibitem{suntajs}
\v{S}untajis J, Bon\v{c}a J, Prosen T and Vidmar L 2019  arXiv:1905.06345

\bibitem{sierant}
Sierant P, DelandeD and Zakrzewski J 2020
{\it Phys. Rev. Lett. }{\bf 124} 186601

\bibitem{abanin} 
Abanin D A, Bardarson J H, De Tomasi G, Gopalakrishnan S, Khemani V, Parameswaran S A,
Pollmann F, Potter A C, Serbyn M and Vasseur R 2019 arXiv:1911.04501

\bibitem{panda}
Panda R K, Scardicchio A, Schulz M, TaylorS R and \v{Z}nidari\v{c} M 2019
{\it Europhys. Lett. }{\bf 128} 67003 

\bibitem{keifer}
Keifer-Emmanouilidis M, Unanyan R, Sirker J and Fleischhauer M 2020
{\it SciPost Phys. } {\bf 8} 083 

\bibitem{unanyan}
Keifer-Emmanouilidis M, Unanyan R, Fleischhauer M and SirkerJ 2020
{\it Phys. Rev. Lett. } {\bf 124} 243601 

\bibitem{max}
Keifer-Emmanouilidis M, Unanyan R, Fleischhauer M and  Sirker J 2020
 arXiv:2010.00565.

\bibitem{Chanda}
Chanda T, Sierant P and Zakrzewski J 2020 {\it Phys. Rev. Research} {\bf 2}, 032045(R) 

\bibitem{Mace}
Mace N, Alet F and Laflorencie N 2019
{\it Phys. Rev. Lett. } {\bf 123} 180601

\bibitem{Yucheng}
Yucheng W, Xiong-junL and DapengY 2019
 arXiv:1910.12080 (2019).

\bibitem{Luitz2} 
Luitz D J,  Khaymovich I M and  Lev Y B 2020
{\it SciPost Phys. Core} {\bf 2} 006 

\bibitem{Makse}
Makse H A, Havlin S, Schwartz M and Stanley H E 1996 
{\it  Phys. Rev. }E {\bf 53} 5445

\bibitem{Janarek}
Janarek J, Delande D and Zakrzewski J 2018
{\it Phys. Rev.} B {\bf 97} 155133

\bibitem{Liu}
Liu F, Ghosh S and Chong Y D 2015
{\it Phys. Rev.} B {\bf 91} 014108 

\bibitem{Mondragon}
IMondragon-Shem I and  Hughes T L 2014
{\it Phys. Rev.} B {\bf 90} 104204 

\bibitem{Mondragon2}
Mondragon-Shem I, Khan m and Hughes T L 2013
{\it Phys. Rev. Lett.} {\bf 110} 046806 

\bibitem{Peschel}
Peschel I 2003
{\it J. Phys. }A. Math. Gen. {\bf 36} 12 

\bibitem{Peschel2}
Peschel I and Eisler V 2009
{\it J. Phys.} A Math. Theor. {\bf 42} 504003 


\bibitem{Calabrese}
Calabrese P and Cardy J 2004
{\it J. Stat. Mech. }{\bf P06002} 

\bibitem{Liu_comp}
This behavior is reminiscent to the previous study \cite{Liu}.

\bibitem{Takaishi}
Takaishi T, Sakakibara K, Ichinose I and Matsui T 2018
{\it  Phys. Rev. }B {\bf 98} 184204 

\bibitem{Hu}
Hu H, Cheng C, Xu Z, Luo H G and Chen S 2014
{\it Phys. Rev.} B {\bf 90} 035150 

\bibitem{Hu2}
Hu H, Chen S, Zeng T S and  Zhang C 2019
{\it Phys. Rev.} A {\bf 100} 023616 

\bibitem{Orito}
Orito T, Kuno Y and Ichinose I 2019 
{\it Phys. Rev. }B {\bf 100} 214202

\bibitem{quspin1}
Weinberg P and Bukov M 2017
{\it SciPost Phys.} {\bf 2} 003 

\bibitem{quspin2}
Weinberg P and Bukov M 2019
{\it SciPost Phys.} {\bf 7} 020 

\bibitem{Page}
Page D N 1993
{\it Phys. Rev. Lett. }{\bf 71} 1291 

\bibitem{Harris}
Harris A B 1974
{\it J. Phys. C: Solid State Phys. }{\bf 7} 1671 

\bibitem{Luitz}
Luitz D J, Laflorencie N and Alet F 2015
{\it Phys. Rev. }B {\bf 91} 081103 

\bibitem{Chandran1}
Chandran A, Pal A, Laumann C R and Scardicchio A 2016
{\it Phys. Rev.} B {\bf 94} 144203 


\bibitem{Serbyn}
Serbyn M, Papic Z and Abanin D A 2013
{\it Phys. Rev. Lett. }{\bf 111} 127201 

\bibitem{Chandran}
Chandran A, Kim I H, Vidal G and Abanin D A 2015
{\it Phys. Rev. }B {\bf 91} 085425

\bibitem{Huse}
Huse D A and Oganesyan V 2014
{\it Phys. Rev. }B {\bf 90} 174202 

\bibitem{Imbrie}
Imbrie J Z, RosV and Scardicchio A 2017
{\it Annalen der Physik} {\bf 529} 1600278 

\bibitem{Bera}
Bera S, Schomerus H, Heidrich-Meisner F and Bardarson J H 2015
{\it Phys. Rev. Lett.} {\bf 115} 046603 

\bibitem{Hopjan}
Hopjan M and Heidrich-Meisner F 2020
{\it Phys. Rev. }A {\bf 101} 063617 






\end{thebibliography}


\end{document}